\documentclass[12pt]{article}
\input{psfig.sty}
\input{epsf.sty}
\newcommand{\Define}{\stackrel{\triangle}{=}}
\usepackage{amssymb}
\newtheorem{thm}{\bf Theorem}
\newtheorem{cor}[thm]{Corollary}
\newtheorem{lem}{{ Lemma}}
\begin{document}

\baselineskip 0.25in

\title{\Large Single-Symbol ML Decodable Distributed STBCs for 
Partially-Coherent Cooperative Networks} 
\author{D. Sreedhar, A. Chockalingam, and B. Sundar Rajan\footnote{This 
work in part was presented in IEEE ICC'2008, Beijing, China, May 2008, and in
IEEE ISIT'2008, Toronto, Canada, July 2008. 
This work was partly supported by the Swarnajayanti Fellowship to 
A. Chockalingam from the Department of Science and Technology, Government 
of India (Project Ref. No: 6/3/2002-S.F.), and by the Council of 
Scientific \& Industrial Research (CSIR), India, through Research Grant 
(22(0365)/04/EMR-II) to B. S. Rajan. This work was also partly supported 
by the DRDO-IISc Program on Advanced Research in Mathematical Engineering 
through research grants to and A. Chockalingam and B. S. Rajan. 
} 
\\
{\normalsize Department of ECE, Indian Institute of Science, 
Bangalore 560012, INDIA}  \\
}
\date{}
\maketitle
\baselineskip 2.00pc
\begin{abstract}
\vspace{-1mm}
Space-time block codes (STBCs) that are single-symbol decodable (SSD) in 
a co-located multiple antenna setting need not be SSD in a distributed 
cooperative communication setting. A relay network with $N$ relays and 
a single source-destination pair is called a partially-coherent relay 
channel (PCRC) if the destination has perfect channel state information 
(CSI) of all the channels and the relays have only the phase information 
of the source-to-relay channels. In this paper, first, a new set of 
necessary and sufficient conditions for a STBC to be SSD for co-located 
multiple antenna communication is obtained. Then, this is extended to a 
set of necessary and sufficient conditions for a distributed STBC (DSTBC) 
to be SSD for a PCRC, by identifying the additional conditions. Using this, 
several SSD DSTBCs for PCRC are identified among the known classes of STBCs. 
It is proved that even if a SSD STBC for a co-located MIMO channel does not 
satisfy the additional conditions for the code to be SSD for a PCRC, 
single-symbol decoding of it in a PCRC gives full-diversity and only 
coding gain is lost. It is shown that when a DSTBC is SSD for a PCRC, 
then arbitrary coordinate interleaving of the in-phase and quadrature-phase
components of the variables does not disturb its SSD property 
for PCRC. Finally, it is shown that the possibility of {\em channel phase 
compensation} operation at the relay nodes using partial CSI at the relays 
increases the possible rate of SSD DSTBCs from $\frac{2}{N}$ when the relays 
do not have CSI to $\frac{1}{2}$, which is independent of $N$.  
\end{abstract}

\vspace{-6mm}
\thispagestyle{empty}
\baselineskip 1.85pc

\newpage

\section{Introduction}
\label{sec1}
\vspace{-3mm}
The problem of fading and the ways to combat it through spatial diversity
techniques have been an active area of research. Multiple-input 
multiple-output (MIMO) techniques have become popular in realizing spatial
diversity and high data rates through the use of multiple transmit antennas. 
For such co-located multiple transmit antenna systems low maximum-likelihood
(ML) decoding complexity space-time block codes (STBCs) have been studied by 
several researchers \cite{TJC}-\cite{KaR}  
which include the well known complex orthogonal designs (CODs) and their 
generalizations. Recent research has shown that the advantages of spatial 
diversity could be realized in single-antenna user nodes through user 
cooperation \cite{SEA},\cite{LaW} via relaying. A simple wireless relay 
network of $N+2$ nodes consists of a single source-destination pair with 
$N$ relays. For such relay channels, use of CODs 
\cite{TJC},\cite{TiH} has been studied in \cite{JiJ}. CODs are attractive 
for cooperative communications for the following reasons: $i)$ they offer 
full diversity gain and coding gain, $ii)$ they are `scale free' in the 
sense that deleting some rows does not affect the orthogonality, $iii)$ 
entries are linear combination of the information symbols and their
conjugates which means only linear  processing is required at the relays,
and $iv)$ they admit very fast ML decoding (single-symbol decoding (SSD)). 
However, it should be noted that the last property applies only to the
decode-and-forward (DF) policy at the relay node. 

In a scenario where the relays amplify and forward (AF) the signal, it is 
known that the orthogonality is lost, and hence the destination has to use 
a complex multi-symbol ML decoding or sphere decoding \cite{JiJ},\cite{RaR1}. 
It should be noted that the AF policy is attractive for two reasons: $i)$ 
the complexity at the relay is greatly reduced, and $ii)$ the restrictions 
on the rate because the relay has to decode is avoided \cite{JiH}. 

In order to avoid the complex ML decoding at the destination, in \cite{RaR2}, 
the authors propose an alternative code design strategy and propose a SSD
code for 2 and 4 relays. For arbitrary number of relays, recently in 
\cite{YiK}, distributed orthogonal STBCs (DOSTBCs) have been studied 
and it is shown that if the destination has the complete channel state 
information (CSI) of all the source-to-relay channels and the 
relay-to-destination channels, then the maximum possible rate is upper 
bounded by $\frac{2}{N}$ complex symbols per channel use for $N$ relays. 
Towards improving the rate of transmission and achieving simultaneously 
both full-diversity as well as SSD at the destination, in this paper, we 
study relay channels with the assumption that the relays have the phase 
information of the source-to-relay channels and the destination has the 
CSI of all the channels. Coding for partially-coherent relay channel 
(PCRC, Section \ref{pcrc_sec}) has been studied in \cite{RaR3}, where a 
sufficient condition for SSD has been presented. 

The contributions of this paper can be summarized as follows: 
\vspace{-4mm}
\begin{itemize}
\item First, a new set of necessary and sufficient conditions for a STBC 
to be SSD for co-located multiple antenna communication is obtained. The 
known set of necessary and sufficient conditions in \cite{KhR} is in terms 
of the dispersion matrices (weight matrices) of the code, whereas our new 
set of conditions is in terms of the column vector representation matrices 
\cite{Lia} of the code and is a generalization of the conditions given in 
\cite{Lia} in terms of column vector representation matrices for CODs.

\item A set of necessary and sufficient conditions for a distributed STBC 
(DSTBC) to be SSD for a PCRC is obtained by identifying the additional 
conditions. Using this, several SSD DSTBCs for PCRC are identified among 
the known classes of STBCs for co-located multiple antenna system. 

\item It is proved that even if a SSD STBC for a co-located MIMO channel 
does not satisfy the additional conditions for the code to be SSD for a 
PCRC, single-symbol decoding of it in a PCRC gives full-diversity and only 
coding gain is lost.  

\item
It is shown that when a DSTBC is SSD for a PCRC, then arbitrary
coordinate interleaving of the in-phase and quadrature-phase components 
of the variables does not disturb its SSD property for PCRC.

\item It is shown that the possibility of {\em channel phase compensation} 
operation at the relay nodes using partial CSI at the relays increases the 
possible rate of SSD DSTBCs from $\frac{2}{N}$ when the relays do not have 
CSI to $\frac{1}{2},$ which is independent of $N$.

\item Extensive simulation results are presented to illustrate the above 
contributions.
\end{itemize}
\vspace{-5mm}
The remaining part of the paper is organized as follows: In Section 
\ref{sec2}, the signal model for a PCRC is developed. Using this model, in 
Section \ref{sec3}, a new set of necessary and sufficient conditions for a 
STBC to be SSD in a co-located MIMO is presented. Several classes of SSD codes 
are discussed and conditions for full-diversity of a subclass of SSD codes 
is obtained. Then, in Section \ref{sec4}, SSD DSTBCs for PCRC are 
characterized by identifying a set of necessary and sufficient conditions. 
It is shown that the SSD property is invariant under coordinate interleaving 
operations which leads to a class of SSD DSTBCs for PCRC. The class of rate 
half CODs obtained from rate one real orthogonal designs (RODs) by stacking 
construction \cite{TJC} is shown to be SSD for PCRC. Also, it is shown that
SSD codes for co-located MIMO, under suboptimal SSD decoder for PCRC offer 
full diversity. Simulation results and discussion constitute Section 
\ref{sec5}. Conclusions and scope for further work are presented in 
Section \ref{sec6}.

\vspace{-4mm}
\section{System Model}
\label{sec2}
\begin{figure}
\begin{center}
\epsfxsize=10.0cm
\epsfbox{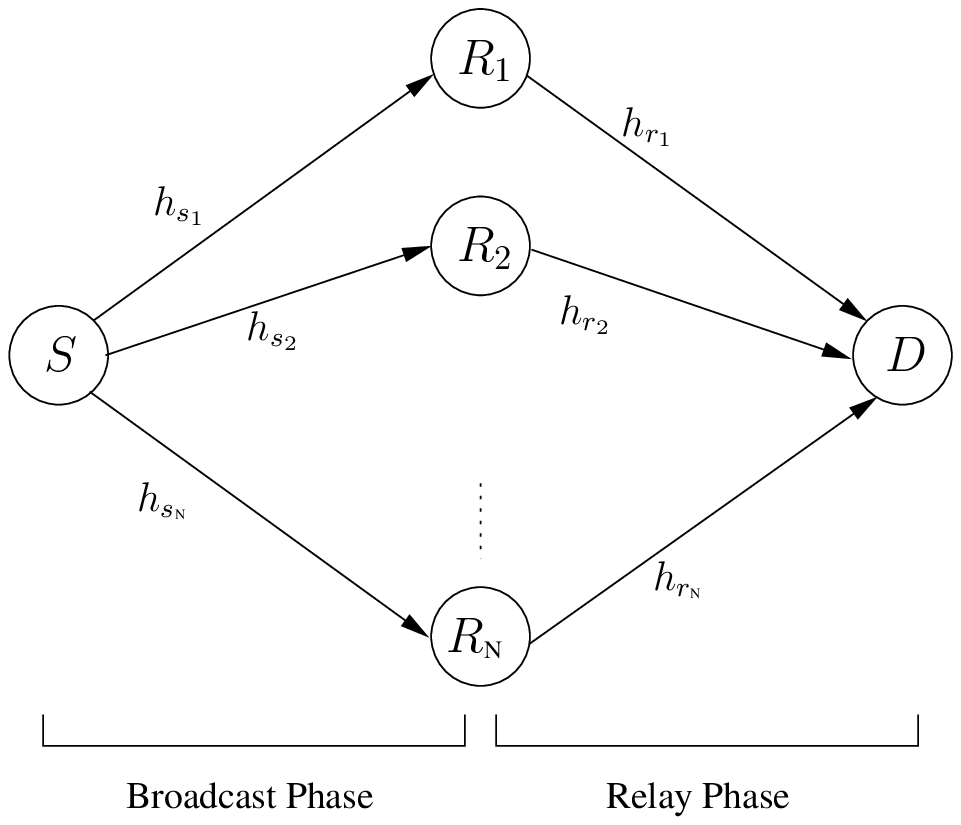}
\caption{A cooperative relay network.}
\vspace{-0mm}
\label{fig1}
\end{center}
\end{figure}

\vspace{-4mm}
Consider a wireless network with $N+2$ nodes consisting of a source, a 
destination, and $N$ relays\footnote{In the system model considered here,
we assume that there is no direct link between source and destination.
However, whatever results we show here can be extended to a system model
with a direct link between source and destination.}, as shown in 
Fig. \ref{fig1}. All nodes are 
half-duplex nodes, i.e., a node can either transmit or receive at a time 
on a specific frequency. We consider amplify-and-forward (AF) transmission 
at the relays. Transmission from the source to the destination is carried 
out in two phases. In the first phase, the source transmits information 
symbols $x^{(i)}, 1 \leq i \leq T_1$ in $T_1$ time slots. All the $N$ relays 
receive these $T_1$ symbols. This phase is called the {\em broadcast phase}. 
In the second phase, all the $N$ relays\footnote{Here, we assume that all 
the $N$ relays participate in the cooperative transmission. It is also 
possible that some relays do not participate in the transmission based 
on whether the channel is in outage or not. We do not consider such a
partial participation scenario here.} perform distributed space-time 
block encoding on their received vectors and transmit the resulting 
encoded vectors in $T_2$ time slots. That is, each relay will transmit
a column (with $T_2$ entries) of a distributed STBC matrix of size
$T_2\times N$. The destination receives a faded and noise added version 
of this matrix. This phase is called the {\em relay phase}. We assume that 
the source-to-relay channels remain static over $T_1$ time slots, and the 
relay-to-destination channels remain static over $T_2$ time slots.

\subsection{No CSI at the relays}
\vspace{-4mm}
The received signal at the $j$th relay, $j=1,\cdots,N$, in the
$i$th time slot, $i=1,\cdots,T_1$, denoted by $v_{j}^{(i)}$, 
can be written as\footnote{We use the following notation: 
Vectors are denoted by boldface lowercase letters, and matrices are 
denoted by boldface uppercase letters. Superscripts $T$ and $\mathcal{H}$ 
denote transpose and conjugate transpose operations, respectively and $*$ 
denotes matrix conjugation operation.}
\begin{eqnarray}
v_{j}^{(i)} & = & \sqrt{E_1} h_{sj} x^{(i)} + z_{j}^{(i)},
\label{rx_relay}
\end{eqnarray}
where $h_{sj}$ is the complex channel gain from the source $s$ to the $j$th 
relay, $z_{j}^{(i)}$ is additive white Gaussian noise at relay $j$ with zero 
mean and unit variance, $E_1$ is the transmit energy per symbol in the 
broadcast phase, and $E\left[\left(x^{(i)}\right)^* x^{(i)}\right] = 1$. 
But no channel knowledge is assumed at the relays. Under the assumption of 
no CSI at the relays, the amplified $i$th received signal at the $j$th relay 
can be written as \cite{JiJ} 
\begin{eqnarray}
\widehat{v}_{j}^{(i)} & = & \underbrace{\sqrt{\frac{E_2}{E_1 + 1}}}_{\Define \,\, G} \, {v}_{j}^{(i)}, 
\label{no_comp}
\end{eqnarray}
where $E_2$ is the transmit energy per transmission of a symbol in the 
relay phase, and $G$ is the amplification factor at the relay that makes 
the total transmission energy per symbol in the relay phase to be equal 
to $E_2$. Let $E_t$ denote the total energy per symbol in both the phases 
put together. Then, it is shown in \cite{JiH} that the optimum energy 
allocation that maximizes the receive SNR at the destination is when half 
the energy is spent in the broadcast phase and the remaining half in the 
relay phase when the time allocations for the relay and broadcast phase are 
same i.e., $T_1 = T_2$. We also assume that the energy is distributed equally 
i.e., $E_1 = \frac{E_t}{2}$ and $E_2 = \frac{E_t}{2M}$, where $M$ is the 
number of transmissions per symbol in the STBC. For the unequal-time 
allocation ($T_1 \neq T_2$) this distribution might not be optimal.

At relay $j$, a $2T_1 \times 1$ real vector $\widehat{\bf v}_{j}$ given by
\begin{eqnarray}
\label{vhatx}
\widehat{\bf v}_{j} & = & \left[ 
\widehat{v}_{jI}^{(1)}, \widehat{v}_{jQ}^{(1)},
\widehat{v}_{jI}^{(2)}, \widehat{v}_{jQ}^{(2)},
\cdots, 
\widehat{v}_{jI}^{(T_1)}, \widehat{v}_{jQ}^{(T_1)}\right]^T,
\end{eqnarray} 
is formed,
where $\widehat{v}_{jI}^{(i)}$ and $\widehat{v}_{jQ}^{(i)}$, 
respectively, are the in-phase (real part) and quadrature (imaginary part) 
components of  $\widehat{v}_{j}^{(i)}.$ Now, (\ref{vhatx}) can be written 
in the form
\begin{eqnarray}
\label{vhat2x}
\widehat{\bf v}_{j} & = & G \, \sqrt{E_1} \, {\bf H}_{sj}\, {\bf x} \, + \, {\bf \widehat{z}}_{j},
\end{eqnarray}
where ${\bf x}$ is the $2T_1 \times 1$ data symbol real vector, given by
\begin{eqnarray}
\label{datavector}
{\bf x} = \left[
x_I^{(1)},x_Q^{(1)},
x_I^{(2)},x_Q^{(2)},
\cdots, 
x_I^{(T_1)}, x_Q^{(T_1)}\right
]^T,
\end{eqnarray}
${\bf \widehat{z}}_{j}$ is the $2T_1 \times 1$ noise vector, given by 
\begin{eqnarray*}
\widehat{\bf z}_{j} & = & \left[
\widehat{z}_{jI}^{(1)}, \widehat{z}_{jQ}^{(1)},
\widehat{z}_{jI}^{(2)}, \widehat{z}_{jQ}^{(2)},
\cdots, 
\widehat{z}_{jI}^{(T_1)}, \widehat{z}_{jQ}^{(T_1)}
\right]^T,
\end{eqnarray*}
where $\widehat{z}_{j}^{(i)} =  G \, z_{j}^{(i)}$, and
${\bf H}_{sj}$ is a $2T_1\times 2T_1$ block-diagonal matrix, given by
\begin{eqnarray}
\label{2bdiagonal}
{\bf H}_{sj} & = & 
\left [ \begin{array}{ccc}
\left[ \begin{array}{cc}
h_{sjI}^{} & -h_{sjQ}  \\ h_{sjQ}  & h_{sjI}
\end{array} \right ]
               & \cdots & {\bf 0} \\
\vdots  & \ddots &  \vdots \\
{\bf 0} & \cdots &
\left[ \begin{array}{cc}
h_{sjI} & -h_{sjQ}  \\ h_{sjQ}  & h_{sjI}
\end{array}\right ]
\end{array}\right ].
\end{eqnarray}
Let 
\begin{eqnarray}
{\bf C} & = & \Big[{\bf c}_1, {\bf c}_2, \cdots, {\bf c}_N\Big]
\end{eqnarray}
denote the $T_2\times N$ distributed STBC matrix to be sent in the 
relay phase jointly by all $N$ relays, where ${\bf c}_j$ denotes
the $j$th column of ${\bf C}$. The $j$th column ${\bf c}_j$ is 
manufactured by the $j$th relay as 
\begin{eqnarray}
\label{stackx}
{\bf c}_{j} & = & {\bf A}_j \widehat{\bf v}_{j} \nonumber \\
& = & \underbrace{G \, \sqrt{E_1} {\bf A}_j \, {\bf H}_{sj}}_{{\bf B}_j}\, {\bf x} \, + \, 
{\bf A}_j \, {\bf \widehat{z}}_{j}, 
\end{eqnarray}
where ${\bf A}_j$ is the complex processing matrix of size $T_2\times 2T_1$ 
for the $j$th relay, called the {\em relay matrix} and ${\bf B}_j$ can be 
viewed as the column vector representation matrix \cite{Lia} for the 
distributed STBC with the difference that in our case the vector ${\bf x}$ 
is real whereas in \cite{Lia} it is complex. For example, for the 2-relay 
case (i.e., $N = 2$), with $T_1=T_2=2,$  using Alamouti code, the relay 
matrices are given by
\begin{eqnarray}
\label{AA}
{\bf A}_1 = \left [ \begin{array}{cccc} 1 & {\bf j} & 0 & 0 \\
0 & 0& -1 & {\bf j}  \end{array} \right ] 
 & \mbox{ and } & 
{\bf A}_2 = \left [ \begin{array}{cccc} 0 &  0 & 1 & {\bf j} \\
1 &  -{\bf j} & 0 & 0 \end{array} \right ].
\end{eqnarray}
Let ${\bf y}$ denote the $T_2 \times 1$ received signal vector at the
destination in $T_2$ time slots. Then, ${\bf y}$ can be written as
\begin{eqnarray}
\label{rx1_no_csi}
{\bf y} & = & \sum_{j=1}^{N} h_{jd} {\bf c}_{j} + {\bf z}_d,
\end{eqnarray}
where $h_{jd}$ is the complex channel gain from the $j$th relay to the 
destination, and ${\bf z}_d$ is the AWGN noise vector at the destination
with zero mean and $E[{\bf z}_d\, {\bf z}_d^*]={\bf I}$. Substituting
(\ref{stackx}) in (\ref{rx1_no_csi}), we can write
\begin{eqnarray}
\label{rx2_no_csi}
{\bf y} & = & G\, \sqrt{E_1} \left(\sum_{j=1}^{N} h_{jd} {\bf H}_{sj} 
{\bf A}_j \right) {\bf x} \, + \, 
\sum_{j=1}^{N} h_{jd} {\bf A}_j {\widehat{\bf z}}_{j} \, + \,{\bf z}_d.
\end{eqnarray}
\subsection{With phase only information at the relays}
\label{pcrc_sec}
\vspace{-3mm}
In this subsection, we obtain a signal model for the case of partial CSI 
at the relays, where we assume that each relay has the knowledge of the 
channel phase on the link between the source and itself in the broadcast 
phase. That is, defining the channel gain from source to relay $j$ as
$h_{sj}=\alpha_{sj}e^{{\bf j}\theta_{sj}}$, we assume that relay $j$
has perfect knowledge of only $\theta_{sj}$ and  does not have the 
knowledge of $\alpha_{sj}.$

\begin{figure}
\begin{center}
\epsfxsize=12.0cm
\epsfbox{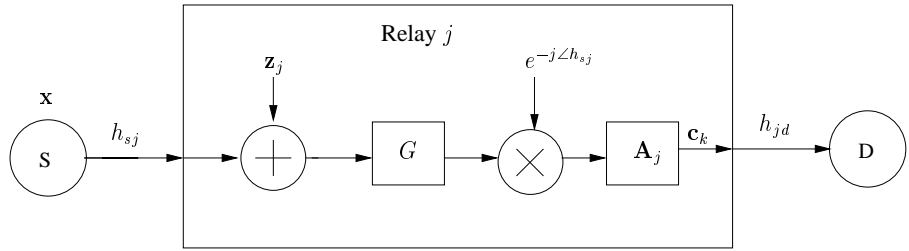}
\vspace{5mm}
\caption{Processing at the $j$th relay in the proposed phase
compensation scheme.}
\label{fig2}
\end{center}
\end{figure}

In the proposed scheme, we perform a phase compensation operation on the
amplified received signals at the relays, and space-time encoding is done
on these phase-compensated signals. That is, we multiply $\widehat{v}_j^{(i)}$
in (\ref{no_comp}) by $e^{-{\bf j}\theta_{sj}}$ before space-time encoding.
Note that multiplication by $e^{-{\bf j}\theta_{sj}}$ does not change the
statistics of ${z}_{j}^{(i)}$. Therefore, with this phase compensation, the
$\widehat{{\bf v}}_j$ vector in (\ref{vhat2x}) becomes
\begin{eqnarray}
\label{vhat2y}
\widehat{\bf v}_{j} & = & \left(G \, \sqrt{E_1} \,  {\bf H}_{sj}\, {\bf x} \, + \, {\bf \widehat{z}}_{j}\right) \, e^{-{\bf j}\theta_{sj}} \nonumber \\
& = & G \, \sqrt{E_1} \, |h_{s_j}| \, {\bf x} \, + \, {\bf \widehat{z}}_{j}.
\end{eqnarray}
Consequently, the ${\bf c}_j$ vector generated by relay $j$ is given by
\begin{eqnarray}
\label{stacky}
{\bf c}_{j} & = & {\bf A}_j \widehat{\bf v}_{j} \nonumber \\
& = & \underbrace{G \, \sqrt{E_1} {\bf A}_j \, \left|h_{sj}\right|}_{\Define \,\,\,{\bf B}_j^{'}}\, {\bf x} \, + \, {\bf A}_j \, {\bf \widehat{z}}_{j},
\end{eqnarray}
where ${\bf B}_j^{'}$ is the equivalent weight matrix with phase compensation.
Now, we can write the received vector $ {\bf y} $ as
\begin{eqnarray}
\label{rx2}
{\bf y} & = & G\, \sqrt{E_1} \left(\sum_{j=1}^{N} h_{jd} |h_{sj}| {\bf A}_j \right) {\bf x} \, + \, \underbrace{\sum_{j=1}^{N} h_{jd} {\bf A}_j {\widehat{\bf z}}_{j} \, + \,{\bf z}_d}_{{\tilde{{\bf z}}}_d \,\,\mbox{{\footnotesize : total noise}}}.
\end{eqnarray}
Figure \ref{fig2} shows the processing at the $j$th relay in the proposed 
phase compensation scheme. Such systems will be referred as {\em
partially-coherent relay channels} (PCRC). A distributed STBC which is 
SSD for a PCRC will be referred as  SSD-DSTBC-PCRC.  
\vspace{-4mm}
\section{Conditions for SSD and Full-Diversity for Co-located MIMO}
\label{sec3}
\vspace{-4mm}
The class of SSD codes, including the well known CODs, for co-located MIMO 
has been studied in \cite{KhR}, where a set of necessary and sufficient 
conditions for an arbitrary linear STBC to be SSD has been obtained in 
terms of the dispersion matrices \cite{HaH}, also known as weight matrices. 
In this section, a new set of necessary and sufficient conditions in terms 
of the column vector representation matrices \cite{Lia} of the code is 
obtained that are amenable for extension to PCRC. This is a generalization 
of the conditions given in \cite{Lia} in terms of column vector 
representation matrices for CODs.
Towards this end, the received vector ${\bf y}$ in a co-located MIMO setup 
can be written as 
\begin{eqnarray}
\label{rx_colocate}
{\bf y} & = & \sqrt{E_t} \left(\sum_{j=1}^{N} h_{jd} {\bf A}_j \right) {\bf x} \, + \,{\bf z}_d.
\end{eqnarray}
\begin{thm}
For co-located MIMO with $N$ transmit antennas, the linear STBC as given
in (\ref{rx_colocate}) is SSD {\em iff }
\begin{eqnarray}
\label{singlesymx}
{\bf A}_{jI}^T{\bf A}_{jI}+ {\bf A}_{jQ}^T{\bf A}_{jQ}&=&{\bf D}_{jj}^{(1)}; ~~~j=1,2,\cdots,N \nonumber \\
{\bf A}_{jI}^T{\bf A}_{iI} +{\bf A}_{jQ}^T{\bf A}_{iQ}+{\bf A}_{iI}^T{\bf A}_{jI}+{\bf A}_{iQ}^T{\bf A}_{jQ}&=&{\bf D}_{ij}^{(2)};~~~ 1 \leq i\neq j \leq N \nonumber \\
{\bf A}_{jI}^T{\bf A}_{iQ} +{\bf A}_{jQ}^T{\bf A}_{iI}-{\bf A}_{iI}^T{\bf A}_{jQ}-{\bf A}_{iQ}^T{\bf A}_{jI}&=&{\bf D}_{ij}^{(3)};~~~ 1 \leq i\neq j\leq N,
\end{eqnarray}
where
${\bf A}_j = {\bf A}_{jI} + {\bf j}{\bf A}_{jQ}, ~ j=1,2,\cdots,N$, where
${\bf A}_{jI}$ and ${\bf A}_{jQ}$ are real matrices, and
${\bf D}_{jj}^{(1)}, {\bf D}_{ij}^{(2)}$ and ${\bf D}_{ij}^{(3)}$  
are block diagonal matrices of the form

\begin{eqnarray}
\label{blockdiagonal}
{\bf D}_{ij}^{(k)} &=&
\left [ \begin{array}{cccc}
\underbrace{
\left[ \begin{array}{cc}
a_{ij,1}^{(k)} & b_{ij,1}^{(k)}  \\ b_{ij,1}^{(k)}  & c_{ij,1}^{(k)}
\end{array} \right ]
}_{{\bf D}_{ij,1}^{(k)}}
             & {\bf 0}   & \cdots & {\bf 0} \\
{\bf 0} &
\underbrace{
\left[ \begin{array}{cc}
a_{ij,2}^{(k)} & b_{ij,2}^{(k)}  \\ b_{ij,2}^{(k)}  & c_{ij,2}^{(k)}
\end{array} \right ]
}_{{\bf D}_{ij,2}^{(k)}}
&  \cdots & {\bf 0} \\
\vdots & \vdots  & \ddots &  \vdots \\
{\bf 0} & \cdots & \cdots & 
\underbrace{
\left[ \begin{array}{cc}
a_{ij,T_1}^{(k)} & b_{ij,T_1}^{(k)}  \\ b_{ij,T_1}^{(k)}  & c_{ij,T_1}^{(k)}
\end{array}\right ]
}_{{\bf D}_{ij,T_1}^{(k)}}
\end{array}\right ],
\end{eqnarray}
where it is understood that whenever the superscript is (1) as in ${\bf D}_{ij}^{(1)},$ then $i=j.$
\end{thm}
{\em Proof:  }
In (\ref{rx2_no_csi}), let 
${\bf H}_{eq} = \sqrt{E_t} \sum_{j=1}^{N} h_{jd} {\bf A}_j $. 
Then the ML optimal detection of ${\bf x}$ is given by 
\begin{eqnarray*}
\widehat{\bf x} &=& \mbox{arg min} \,\, 
|| {\bf y} - {\bf H}_{eq} {\bf x} ||^2.
\end{eqnarray*}
Since ${\bf x}$ is real,
\begin{eqnarray*}
|| {\bf y} - {\bf H}_{eq} {\bf x} ||^2 & = & || {\bf y} ||^2 
- 2 {\bf x}^T \Re \left ( {\bf H}_{eq}^{\mathcal{H}} {\bf y} \right ) +  
 {\bf x}^T \Re \left ( {\bf H}_{eq}^{\mathcal{H}} {\bf H}_{eq} \right ) {\bf x},
\end{eqnarray*}
which can be written as the sum of several metrics each depending only on 
one symbol {\em iff} 
$\Re\left ( {\bf H}_{eq}^{\mathcal{H}} {\bf H}_{eq} \right )$ is a block 
diagonal matrix of the form in (\ref{blockdiagonal}) for every possible 
realization of $h_{jd}$.  Now,
\begin{eqnarray*}
\Re\left ( {\bf H}_{eq}^{\mathcal{H}} {\bf H}_{eq} \right ) &=&
E_t\sum_{j=1}^{N} |h_{jd}|^2 \Re \left ({\bf A}_j^{\mathcal{H}} {\bf A}_j \right ) + 
\nonumber \\
& &
E_t\sum_{j_1=1}^{N} \sum_{j_2=1,j_2 \neq j_1}^{N}  \Re \left (
h^*_{j_1d} h_{j_2d} {\bf A}^{\mathcal{H}}_{j_1} {\bf A}_{j_2} +  
h^*_{j_2d} h_{j_1d} {\bf A}^{\mathcal{H}}_{j_2} {\bf A}_{j_1}
\right ) \nonumber \\
&=&
E_t\sum_{j=1}^{N} |h_{jd}|^2 \Re \left ({\bf A}_j^{\mathcal{H}} {\bf A}_j \right ) + 
\nonumber \\
& &
E_t\sum_{j_1=1}^{N} \sum_{j_2=1,j_2 \neq j_1}^{N} 
\left (h_{j_1dI} h_{j_2dI} + h_{j_1dQ} h_{j_2dQ} \right ) 
\Re \left ({\bf A}^{\mathcal{H}}_{j_1} {\bf A}_{j_2} + {\bf A}^{\mathcal{H}}_{j_2} {\bf A}_{j_1} 
\right ) + 
\nonumber \\
& &
E_t\sum_{j_1=1}^{N} \sum_{j_2=1,j_2 \neq j_1}^{N} 
\left (h_{j_1dI} h_{j_2dQ} - h_{j_1dQ} h_{j_2dI} \right ) 
\Im \left ({\bf A}^{\mathcal{H}}_{j_1} {\bf A}_{j_2} - {\bf A}^{\mathcal{H}}_{j_2} {\bf A}_{j_1}  
\right ),
\end{eqnarray*}
which is block diagonal of the form in (\ref{blockdiagonal}) $\forall h_{jd}$ 
{\em iff} (\ref{singlesymx}) is satisfied\footnote{We note that, for the 
co-located case, SSD conditions have been presented in \cite{KhR} in terms 
of the linear dispersion matrices (also called weight matrices). Our SSD 
conditions given in Theorem 1 is in terms of `column vector representation 
matrices' \cite{Lia}. The significance of our version as in Theorem 1 is 
that it is instrumental in proving Theorems 2 to 6.}. $\square$  

Notice that ${\bf D}_{ij}^{(k)}= {\bf D}_{ji}^{(k)}$ for all $i,j,k.$ The 
conditions for achieving maximum diversity depend on the ${\bf D}_{ij}^{(k)}$ 
matrices as well as the signal constellation used for the variables. Before 
we discuss these conditions in Lemma \ref{lem1}, we illustrate the SSD 
conditions (\ref{singlesymx}) for the following classes of SSD codes for 
co-located MIMO.
\vspace{-3mm}
\subsection{SSD conditions for some known classes of codes}
\vspace{-3mm}
{\bf Complex Orthogonal designs (COD):} STBCs from CODs have been 
extensively studied \cite{TJC},\cite{TiH},\cite{Lia}. A {\textit{Square 
Complex Orthogonal Design}} (SCOD) $ {\bf G} (x_1,x_2,\cdots,x_K)$  
(in short ${\bf G}$) of size $N$ is an  $ N \times N $ matrix such that 
$i)$ the entries of $ {\bf G} (x_1,x_2,\cdots,x_K)$ are complex linear 
combinations of the variables $x_1,x_2,\cdots,x_K $ and their complex 
conjugates $x_1^*,x_2^*,\cdots,x_K^*$, and  
${\bf G}^{\mathcal{H}} {\bf G}=({\vert x_1\vert}^2 +\cdots+{\vert x_K\vert}^2) {\bf I}_N$,
where ${\bf I}_N$ is the $ N \times N $ identity matrix. The rate of 
${\bf G}$ is $\frac{K}{N}$ complex symbols per channel use. SCODs 
$COD_{2^a}$ for $2^a$ antennas, $a=2,3,\cdots$, can be recursively 
constructed starting from
\begin{equation}
\label{itcod}
COD_2= \left[\begin{array}{rr}
x_1   &-x_2^*      \\
x_2   & x_1^*
\end{array}\right],~~~
COD_{2^a}=  \left[\begin{array}{rr}
{\bf G}_{a-1}   & -x_{a+1}^*{\bf I}_{2^{a-1}}      \\
x_{a+1}{\bf I}_{2^{a-1}}   & {\bf G}_{a-1}^{\mathcal{H}}
\end{array}\right],
\end{equation}
\noindent
where $G_{2^a}$ is a $2^a\times 2^a$ complex matrix. 
For example, 
\begin{eqnarray}
\label{itcod4}
COD_4 &=& \left [\begin{array} {cccc} 
x_1  & x_2 & -x_3^* & 0 \\
-x_2^* & x_1^* & 0 & -x_3^* \\
x_3 & 0 & x_1^* & -x_2 \\
0 & x_3 & x_2^* & x_1
\end{array} \right ], 
\end{eqnarray}
\begin{eqnarray}
\label{itcod8}
COD_8 &=& \left [\begin{array} {cccccccc}
x_1    & x_2   & -x_3^* & 0     & -x_4^* & 0     & 0      & 0\\
-x_2^* & x_1^* & 0      &-x_3^* & 0      & -x_4^*& 0      & 0 \\
x_3    & 0     & x_1^*  &-x_2   & 0      & 0     & -x_4^* & 0 \\
0      & x_3   & x_2^* & x_1    & 0      & 0     & 0      & -x_4^* \\
x_4      & 0     & 0      & 0     & x_1^*  & -x_2  & x_3^*  & 0 \\
0      & x_4     & 0      & 0     & x_2^*  & x_1   & 0      &x_3^* \\
0      & 0     & x_4      & 0     & -x_3   &  0    & x_1    & x_2  \\
0      & 0     & 0      & x_4     & 0      & -x_3  & -x_2^* & x_1^*
\end{array} \right ].
\end{eqnarray}
Any COD, ${\bf G}$, can be written as
\begin{eqnarray}
{\bf G} = \left [ {\bf A}_1{\bf x},{\bf A}_2{\bf x},\cdots,{\bf A}_N {\bf x} \right ],
\end{eqnarray}
where $ {\bf A}_1, {\bf A}_2,\cdots, {\bf A}_N$ are the relay matrices.
By the definition of CODs, ${\bf G}^{\mathcal{H}} {\bf G} = 
\left ( {\bf x}^T {\bf x} \right ) {\bf I}$, 
which implies that 
\begin{eqnarray}
\label{codd1}
{\bf x} ^T {\bf A}_j^{\mathcal{H}}  {\bf A}_j {\bf x} &= & 
{\bf x}^T {\bf x};~~~ \forall ~~~ j \\
\label{codd2}
{\bf x} ^T {\bf A}_j^{\mathcal{H}}  {\bf A}_i {\bf x} &= & 0; 
~~~ \forall ~~~ i\neq j.
\end{eqnarray}
Eqn. (\ref{codd1}) implies that $ \Re \left ( {\bf A}_j^{\mathcal{H}}  
{\bf A}_j \right ) = {\bf I}\,\,\forall j$, i.e., 
$ {\bf D}^{(1)}_{jj} = {\bf I} \,\, \forall j$,
whereas Eqn. (\ref{codd2}) implies that
\begin{eqnarray}
 \left ( {\bf A}_j^{\mathcal{H}}  {\bf A}_i \right )^T &=& 
- {\bf A}_j^{\mathcal{H}}  {\bf A}_i;~~~ \forall ~~ i\neq j,
\end{eqnarray}
which implies that 
\begin{eqnarray}
{\bf A}_{jI}^T{\bf A}_{iI} +{\bf A}_{jQ}^T{\bf A}_{iQ}+{\bf A}_{iI}^T{\bf A}_{jI}+{\bf A}_{iQ}^T{\bf A}_{jQ} & = & {\bf D}_{ij}^{(2)} 
 =  {\bf 0};
~~~ \forall ~~~ i\neq j 
\nonumber \\
{\bf A}_{jI}^T{\bf A}_{iQ} +{\bf A}_{jQ}^T{\bf A}_{iI}-{\bf A}_{iI}^T{\bf A}_{jQ}-{\bf A}_{iQ}^T{\bf A}_{jI} & = & {\bf D}_{ij}^{(3)} 
 =  {\bf 0};
~~~ \forall ~~~ i\neq j.
\end{eqnarray}
Hence, for CODs $ {\bf D}^{(2)}_{ij} = {\bf D}^{(3)}_{ij} = {\bf 0} \,\, 
\forall \,\, i,j $ and ${\bf D}_{jj}^{(1)}$ is the identity matrix $ \forall \,\, j$. 

{\bf Coordinate Interleaved Orthogonal designs (CIOD) \cite{KhR}:}

A coordinate interleaved orthogonal design (CIOD) 
in variables $x_{i}, i=0,\cdots, K-1$ (where $K$ is even) is a $2L \times 2N$
matrix ${\bf S}$, such that
\begin{equation}
\label{c2eq1}
{\bf S}({x}_{0},\cdots,{x}_{K-1})=\left[\begin{array}{cc} \Theta(\tilde{x}_0,\cdots,\tilde{x}_{K/2-1}) &0\\
0 & \Theta(\tilde{x}_{K/2},\cdots,\tilde{x}_{K-1})
\end{array}\right],
\end{equation}
where $\Theta(x_0,\cdots,x_{K/2-1})$ is generalized COD (GCOD) of size 
$L \times N$ and rate $K/2L$ and 
$\tilde{x}_i=\Re (x_i)+ {\bf j} \Im(x_{({i+K/2})_K})$ and 
$(a)_K$ denotes $(a \,\, \mbox{mod} \,\, K)$.
Consider the four transmit antenna CIOD, denoted by
$CIOD_4$:
\begin{eqnarray}
\label{ciod4}
CIOD_4 &=& \left [\begin{array} {cccc}
\tilde{x}_0  & \tilde{x}_1 & 0 & 0 \\
-\tilde{x_1}^* & \tilde{x_0}^* & 0 & 0 \\
0 & 0 & \tilde{x}_2 & \tilde{x}_3 \\
0 & 0 & -\tilde{x}_3^* & \tilde{x}_2^*
\end{array} \right ],
\end{eqnarray}
where $ \tilde{x}_i = x_{iI} + {\bf j} x_{{((i + 2)~mod~4})Q}$, and the eight 
transmit antenna CIOD, denoted by $CIOD_8$:
\begin{eqnarray}
\label{ciod8}
CIOD_8 &=& \left [\begin{array} {cccccccc}
\tilde{x}_1  & \tilde{x}_2 &\tilde{x}_3 & 0 & 0 & 0& 0 & 0\\
-\tilde{x}_2^* &\tilde{x}_1^* & 0 &\tilde{x}_3  & 0 & 0& 0 & 0 \\
-\tilde{x}_3^* & 0 & \tilde{x}_1 &\tilde{x}_2 & 0 & 0& 0 & 0 \\
0 & -\tilde{x}_3^* & -\tilde{x}_2^* & \tilde{x}_1^*  & 0 & 0& 0 & 0 \\
0 & 0& 0 & 0 & \tilde{x}_4  & \tilde{x}_5 & \tilde{x}_6 & 0 \\
0 & 0& 0 & 0 & -\tilde{x}_5^* & \tilde{x}_4^* & 0 &\tilde{x}_6 \\
0 & 0& 0 & 0 & -\tilde{x}_6^* & 0 & \tilde{x}_4 & \tilde{x}_5  \\
0 & 0& 0 & 0 & 0 & -\tilde{x}_6^* & -\tilde{x}_5^* & \tilde{x}_4^*
\end{array} \right],
\end{eqnarray}
where $ \tilde{x}_i = x_{iI} + {\bf j} x_{{((i + 4)~mod~4})Q}.$ 
The data-symbol vector in (\ref{datavector}) after interleaving can be 
written as
\begin{eqnarray}
\tilde{\bf x} & = & \tilde{\bf I} \, {\bf x},
\end{eqnarray}
where $\tilde{\bf I}$ is the interleaving matrix, which is a permutation 
matrix obtained by permuting the rows (/columns) of the identity matrix 
{\bf I} to reflect the coordinate interleaving operation. Hence, the 
effective relay matrices of the design ${\bf S}, \, \bar{\bf A}_j $, 
can be written as
$ \bar{\bf A}_j =
\left [ \begin{array} {cc} {\bf A}_{j} & {\bf 0}_{L \times K} \\ 
{\bf 0}_{L \times K} & {\bf 0}_{L \times K}  
\end{array} \right ] 
\tilde{\bf I},\,\, 1 \leq j \leq N $ and
$ \bar{\bf A}_j =
\left [ \begin{array} {cc}  {\bf 0}_{L \times K} & {\bf 0}_{L \times K} \\  
{\bf 0}_{L \times K} & {\bf A}_{j} \end{array} \right ]
\tilde{\bf I} ,\,\, N + 1 \leq j \leq 2N $, where ${\bf A}_{j}$'s
are relay matrices of the design $ \Theta$. It can be verified that 
$ {\bf D}_{jj}^{(1)} =  \tilde{\bf I}^T
\left [ \begin{array} {cc} {\bf I}_{K \times K} & {\bf 0}_{K \times K} \\ 
{\bf 0}_{K \times K} & {\bf 0}_{K \times K}  
\end{array} \right ]   \tilde{\bf I} $ for $ 1 \leq j \leq N $ and 
$ {\bf D}_{jj}^{(1)} =  \tilde{\bf I}^T
\left [ \begin{array} {cc} {\bf 0}_{K \times K} & {\bf 0}_{K \times K} \\ 
{\bf 0}_{K \times K} & {\bf I}_{K \times K}  
\end{array} \right ]   \tilde{\bf I} $ for $ N+1 \leq j \leq 2N$.
Also, ${\bf D}_{ij}^{(2)} = {\bf D}_{ij}^{(3)} = {\bf 0};\,\, 
\forall \,i \neq j $. 
Hence, $ {\bf D}^{(2)}_{ij} = {\bf D}^{(3)}_{ij} = {\bf 0}\, \forall \,\, i,j$ 
for CIODs also. But,  ${\bf D}_{jj}^{(1)}$ is not the identity matrix 
$\forall \,\, j.$ 

\vspace{-0mm}
{\bf Clifford UW-SSD codes \cite{KaR}:}

A $2^a-$Clifford Unitary Weight SSD (CUW-SSD) code, denoted by $CUW_{2^a},$ 
is a $2^a\times 2^a$ STBC, given by 

{\footnotesize
\begin{equation}
\mbox{\hspace{-7mm}}
\begin{array}{l}
 \sigma_{x_1} \bigotimes {\bf I}_2^{\otimes^{a-1}} +\rho_{x_{2a}} \bigotimes \sigma_3 ^{\otimes^{a-1}}
  + \sum_{i=1}^{a-1}
\left[
\sigma_{x_{2i}} \bigotimes {\bf I}_2^{\otimes^{a-i-1}}  \bigotimes \sigma_1 \bigotimes \sigma_3^{\otimes^{i-1}}+\sigma_{x_{2i+1}} \bigotimes {\bf I}_2^{\otimes^{a-i-1}} \bigotimes \sigma_2 \bigotimes \sigma_3^{\otimes^{i-1}}
\right], \mbox{\hspace{-3mm}}
\end{array}
\label{cuwssdeq}
\end{equation} 
}

\vspace{-8mm}
where
\[\begin{array}{c}
x_i = x_{iI}+jx_{iQ},~~
\sigma_{x_i} = \left[
\begin{array}{rr}
x_{iI} & jx_{iQ} \\
-jx_{iQ} & x_{iI}
\end{array}
\right], ~~ 
\rho_{x_i} = \left[
\begin{array}{rr}
-jx_{iQ} & jx_{iI} \\
-x_{iI} & -jx_{iQ}
\end{array}
\right],
\end{array}
\]
\begin{equation}
\label{paulimatrices}
\sigma_1 =\left[ \begin{array}{rr}
0 & 1 \\
-1 & 0
\end{array}
\right],~~
\sigma_2 =\left[ \begin{array}{rr}
0 & j \\
j & 0
\end{array}
\right],~~
\sigma_3 =\left[ \begin{array}{rr}
1 & 0 \\
0 & -1
\end{array}
\right],
\end{equation}
and 
$\bigotimes$ stands for the tensor product of matrices. 
Based on the above definition, the $2-$CUW-SSD code is given by
\begin{equation}
\begin{array}{lll}
CUW_2& = & \sigma_{x_1}+\rho_{x_2} = \left[
\begin{array}{rr}
x_{1I}-jx_{2Q}  & x_{2I}+jx_{1Q} \\
-x_{2I}-jx_{1Q} & x_{1I}-jx_{2Q}
\end{array}
\right],
\end{array}
\end{equation}
and the $4-$CUW-SSD code is given by
\begin{eqnarray}
\begin{array}{lll}
CUW_4& = &\sigma_{x_1}\bigotimes {\bf I}_2 + \rho_{x_1} \bigotimes \sigma_3 +\sigma_{x_2} \bigotimes \sigma_1 +\sigma_{x_3}\bigotimes \sigma_2,
\end{array}
\end{eqnarray}
which is
\begin{eqnarray}
\label{cliff4}
CUW_4 &=& \left [ \begin{array} {cccc}
 x_{1I} - {\bf j} x_{4Q} &  x_{2I} + {\bf j} x_{3I} &
 x_{4I} + {\bf j} x_{1Q} & -x_{3Q} + {\bf j} x_{2Q} \\
-x_{2I} - {\bf j} x_{3I} &  x_{1I} + {\bf j} x_{4Q} &
-x_{3Q} - {\bf j} x_{2Q} & -x_{4I} + {\bf j} x_{1Q} \\
-x_{4I} - {\bf j} x_{1Q} &  x_{3Q} - {\bf j} x_{2Q} &
 x_{1I} - {\bf j} x_{4Q} &  x_{2I} + {\bf j} x_{3I} \\
 x_{3Q} + {\bf j} x_{2Q} &  x_{4Q} - {\bf j} x_{1Q} &
-x_{2I} - {\bf j} x_{3I} &  x_{1I} + {\bf j} x_{4Q}
\end{array} \right ].
\end{eqnarray}
It can be verified that for Clifford UW-SSD codes 
${\bf D}^{(2)}_{ij} = {\bf 0} \,  \forall \,i,j$, and the matrices   
${\bf D}^{(3)}_{ij}\, \forall \,i,j$ and ${\bf D}_{jj}^{(1)} \, \forall \, j$ 
are of the form (\ref{blockdiagonal}). For example, for the $CUW_2$ code, 
${\bf D}^{(1)}_{jj} = {\bf I} \,\, \forall j$, 
${\bf D}^{(2)}_{ij} = {\bf 0} \,\, \forall i,j$, 
\begin{footnotesize}
\begin{eqnarray}
{\bf D}^{(3)}_{1,2} = -{\bf D}^{(3)}_{2,1} = 
\left [ 
\begin{array} {cccc}
0 & 2 & 0 & 0  \\
2 & 0 & 0 & 0  \\
0 & 0 & 0 & 2  \\
0 & 0 & 2 & 0  \\
\end{array}
\right ], 
\end{eqnarray}
\end{footnotesize}
and ${\bf D}^{(3)}_{ij} = {\bf 0} $ for all other values of $i,j$. 
For the $ CUW_4 $ code, $ {\bf D}^{(1)}_{jj} = {\bf I} \,\, 
\forall j $, $ {\bf D}^{(2)}_{ij} = {\bf 0} \,\, \forall i,j$, 
\begin{footnotesize}
\begin{eqnarray}
{\bf D}^{(3)}_{1,3} = {\bf D}^{(3)}_{2,4} = -{\bf D}^{(3)}_{3,1} = 
-{\bf D}^{(3)}_{4,2} =  
\left [ 
\begin{array} {cccccccc}
0 & 2 & 0 & 0 & 0 & 0 & 0 & 0 \\
2 & 0 & 0 & 0 & 0 & 0 & 0 & 0 \\
0 & 0 & 0 & 2 & 0 & 0 & 0 & 0 \\
0 & 0 & 2 & 0 & 0 & 0 & 0 & 0 \\
0 & 0 & 0 & 0 & 0 & 2 & 0 & 0 \\
0 & 0 & 0 & 0 & 2 & 0 & 0 & 0 \\
0 & 0 & 0 & 0 & 0 & 0 & 0 & 2 \\
0 & 0 & 0 & 0 & 0 & 0 & 2 & 0 \\
\end{array}
\right ], 
\end{eqnarray}
\end{footnotesize}
and $ {\bf D}^{(3)}_{ij} = {\bf 0} $ for all other values of $i,j$. 

\vspace{-3mm}
\subsection{Conditions for full-diversity}
\vspace{-3mm}
In the previous subsection, we saw several classes of SSD codes. The 
problem of identifying all possible classes of SSD codes is a open 
problem \cite{KaR}. Moreover, different classes of SSD codes may give 
full-diversity for different sets of signal sets. The following lemma 
obtains a set of necessary and sufficient conditions for the subclass 
of SSD codes characterized by  
${\bf D}_{ij}^{(2)} = {\bf D}_{ij}^{(3)} = {\bf 0}$ (CODs and CIODs,
for example) to offer full-diversity for all complex constellations. 
\begin{lem}
\label{lem1}
For co-located MIMO, the linear  STBC as given in (\ref{rx_colocate}) 
with the ${\bf D}_{ij}^{(k)}$ matrices in (\ref{singlesymx}) satisfying 
${\bf D}_{ij}^{(2)} = {\bf D}_{ij}^{(3)} = {\bf 0}$ achieves maximum 
diversity for all signal constellations {\em iff} 
\begin{eqnarray}
\label{divcondn}
a^{(1)}_{jj,l} c^{(1)}_{jj,l}  - {b^{(1)}_{jj,l}}^2 > 0, \,\,\,\, 1 \leq j \leq N;~~~1 \leq l \leq T_1,
\end{eqnarray}
i.e., ${\bf D}_{jj,l}^{(1)}$  is positive definite for all $j,l.$ 
\end{lem}
{\em Proof:  } Consider the pairwise error probability that the 
data vector ${\bf x}_1$ as in (\ref{datavector}) gets wrongly detected as
${\bf x}_2$. By Chernoff bound,
\begin{eqnarray}
\label{cb1}
P\left( {\bf x}_1 \rightarrow {\bf x}_2 \right ) \leq 
E \left\{ e^{-d^2({\bf x}_1,{\bf x}_2) E_t/4 } \right\},
\end{eqnarray}
where, from (\ref{rx_colocate}), 
\begin{eqnarray}
\label{cb2}
d^2({\bf x}_1,{\bf x}_2) = 
({\bf x}_2 - {\bf x}_1)^T \Re \left ( {\bf H}_{eq}^{\mathcal{H}} {\bf H}_{eq} \right )({\bf x}_2 - {\bf x}_1).
\end{eqnarray}
Define $ \Delta {\bf x}^{(i)} = [ \Delta x_I^{(i)} \Delta x_Q^{(i)}]^T = 
[  (x_{2I}^{(i)} - x_{1I}^{(j)}), (x_{2Q}^{(i)} - x_{1Q}^{(i)}) ]^T $.
Given that the conditions (\ref{singlesymx}) are satisfied, the distance
metric can be written as sum of $T_1$ terms as
\begin{eqnarray}
d^2({\bf x}_1,{\bf x}_2) &= &
 \sum_{l=1}^{T_1} \Delta {{\bf x}^{(l)}}^T \left ( 
\sum_{j=1}^{N} |h_{jd}|^2 {\bf D}^{(1)}_{jj,l}
\right ) \Delta {\bf x}^{(l)} \nonumber \\
& = &
 \sum_{j=1}^{N} |h_{jd}|^2  \left (
\sum_{l=1}^{T_1} \Delta {{\bf x}^{(l)}}^T  
{\bf D}^{(1)}_{jj,l}  \Delta {\bf x}^{(l)}
\right ).
\label{eq_here}
\end{eqnarray}
Substituting (\ref{eq_here}) in (\ref{cb1}) and evaluating the expectation, 
we obtain
\begin{eqnarray}
\label{cb4}
P\left( {\bf x}_1 \rightarrow {\bf x}_2 \right ) \leq 
\prod_{j=1}^N\left ( \frac{1}{1 + 
\sum_{l=1}^{T_1} \Delta {{\bf x}^{(l)}}^T  
{\bf D}^{(1)}_{jj,l}  \Delta {\bf x}^{(l)} E_t/4
} \right ),
\end{eqnarray}
which, for high SNRs, can be written as
\begin{eqnarray}
\label{cb40}
P\left( {\bf x}_1 \rightarrow {\bf x}_2 \right ) \leq
\prod_{j=1}^N\left ( \frac{1}{
\sum_{l=1}^{T_1} \Delta {{\bf x}^{(l)}}^T
{\bf D}^{(1)}_{jj,l}  \Delta {\bf x}^{(l)} E_t/4
} \right ).
\end{eqnarray}
Hence, for high SNRS, in order to achieve full diversity, 
$\Delta {{\bf x}^{(l)}}^T  {\bf D}^{(1)}_{jj,l}  \Delta {\bf x}^{(l)}$  
should be non-zero for all $j,l$, i.e., ${\bf D}^{(1)}_{jj,l}$ should be 
a positive definite matrix $\forall j,l $, i.e., 
$ a^{(1)}_{j,l} c^{(1)}_{j,l} - {b^{(1)}_{j,l}}^2 > 0 \,\,\forall \,l,j.$ 
$\square$ 

For CODs, by definition, $b^{(1)}_{jj,l} = 0 $ and 
$a^{(1)}_{jj,l} = c^{(1)}_{jj,i} = 1 \forall j,i $. Hence, the condition 
in (\ref{divcondn}) is readily satisfied, and hence full diversity is 
achieved for all signal constellations. However, for CIODs, the condition 
(\ref{divcondn}) is not satisfied as shown below for the code $CIOD_4.$ 
For this code,

\begin{footnotesize}
\begin{eqnarray}
{\bf A}_{1I}^T{\bf A}_{1I}+ {\bf A}_{1Q}^T{\bf A}_{1Q} \,\,\, = \,\,\,
{\bf A}_{2I}^T{\bf A}_{2I}+ {\bf A}_{2Q}^T{\bf A}_{2Q} & = &
\left [ 
\begin{array} {cccccccc}
1 & 0 & 0 & 0 & 0 & 0 & 0 & 0 \\
0 & 0 & 0 & 0 & 0 & 0 & 0 & 0 \\
0 & 0 & 1 & 0 & 0 & 0 & 0 & 0 \\
0 & 0 & 0 & 0 & 0 & 0 & 0 & 0 \\
0 & 0 & 0 & 0 & 0 & 0 & 0 & 0 \\
0 & 0 & 0 & 0 & 0 & 1 & 0 & 0 \\
0 & 0 & 0 & 0 & 0 & 0 & 0 & 0 \\
0 & 0 & 0 & 0 & 0 & 0 & 0 & 1 \\
\end{array}
\right ];  \\
{\bf A}_{3I}^T{\bf A}_{3I}+ {\bf A}_{3Q}^T{\bf A}_{3Q} \,\,\, = \,\,\,
{\bf A}_{4I}^T{\bf A}_{4I}+ {\bf A}_{4Q}^T{\bf A}_{4Q} & = &
\left [ 
\begin{array} {cccccccc}
0 & 0 & 0 & 0 & 0 & 0 & 0 & 0 \\
0 & 1 & 0 & 0 & 0 & 0 & 0 & 0 \\
0 & 0 & 0 & 0 & 0 & 0 & 0 & 0 \\
0 & 0 & 0 & 1 & 0 & 0 & 0 & 0 \\
0 & 0 & 0 & 0 & 1 & 0 & 0 & 0 \\
0 & 0 & 0 & 0 & 0 & 0 & 0 & 0 \\
0 & 0 & 0 & 0 & 0 & 0 & 1 & 0 \\
0 & 0 & 0 & 0 & 0 & 0 & 0 & 0 \\
\end{array}
\right ]. 
\end{eqnarray}
\end{footnotesize}

\vspace{-4mm}
Hence, none of the ${\bf D}^{(1)}_{jj} $ matrices are positive definite. 
This does not mean that the code can not give full diversity; it only 
means that it can not give full diversity for all complex constellations 
as mentioned in Lemma \ref{lem1}. The constellations for which this code 
offers full diversity can be obtained by choosing the signal constellation 
such that for any two constellation points, $\Delta {{\bf x}_I^{(i)}}$ and 
$\Delta {{\bf x}_Q^{(i)}}$  are both non-zero. Substituting these values
in the pair-wise error probability expression (\ref{cb4}), we get
\begin{eqnarray}
\label{cb5}
P\left( {\bf x}_1 \rightarrow {\bf x}_2 \right ) \leq
\prod_{i=1}^{2} 
\left ( \frac{1}{1 + \Delta {{\bf x}_I^{(i)}}^2 E_t/4} \right )
\left ( \frac{1}{1 + \Delta {{\bf x}_Q^{(i)}}^2 E_t/4} \right ).
\end{eqnarray}
This has already been shown in \cite{KhR}. 

\vspace{-4mm}
\section{SSD Codes for PCRC}
\label{sec4}
\vspace{-5mm}
In the previous section, we saw that SSD is achieved if the relay matrices 
satisfy the condition (\ref{singlesymx}). However, to achieve SSD in the 
case of distributed STBC with AF protocol, the equivalent weight matrices 
${\bf B}_j$'s must satisfy the condition in (\ref{singlesymx}). It can be 
seen that for any ${\bf A}_j $ that satisfies the condition in 
(\ref{singlesymx}), the corresponding  ${\bf B}_j$'s need not satisfy 
(\ref{singlesymx}). For example, for the weight matrices in (\ref{AA}), 
the corresponding equivalent weight matrices ${\bf B}_1$ and ${\bf B}_2$ 
do not satisfy the condition in (\ref{singlesymx}). That is, the Alamouti 
code is not SSD as a distributed STBC with AF protocol. We note that, in 
\cite{RaR2}, code designs which retain the SSD feature have been obtained 
for no CSI at the relays, but only for $N=2$ and $4$. A key contribution 
in this paper is that by using partial CSI at the relays (i.e., only the 
channel phase information of the source-to-relay links), the SSD feature 
at the destination can be restored for a large subclass of SSD codes for 
co-located MIMO communication. 
This key result is given in the following theorem, which 
characterizes the class of SSD codes for PCRC.
\begin{thm} 
\label{thm2}
A code as given by (\ref{stackx}) is SSD-DSTBC-PCRC {\em iff } the  relay 
matrices ${\bf A}_j,~ j=1,2,\cdots,N,$ satisfy (\ref{singlesymx}) (i.e., 
the code is SSD for a co-located MIMO set up), and, in addition, 
for any three relays with indices $j_1$, $j_2$, $j_3$, where
$j_1,j_2,j_3 \in \{1,2,\cdots,N\}$,
{\small
\begin{eqnarray}
\label{ssddstbc1}
{{\bf A}_{j_1I}}^T\left( 
{{\bf A}_{j_2I}} {{\bf A}_{j_2I}}^T + 
{{\bf A}_{j_2Q}} {{\bf A}_{j_2Q}}^T
\right ){{\bf A}_{j_3I}} + 
{{\bf A}_{j_3I}}^T\left( 
{{\bf A}_{j_2I}} {{\bf A}_{j_2I}}^T + 
{{\bf A}_{j_2Q}} {{\bf A}_{j_2Q}}^T
\right ){{\bf A}_{j_1I}} + & &\nonumber \\
{{\bf A}_{j_1Q}}^T\left( 
{{\bf A}_{j_2I}} {{\bf A}_{j_2I}}^T + 
{{\bf A}_{j_2Q}} {{\bf A}_{j_2Q}}^T
\right ){{\bf A}_{j_3Q}} + 
{{\bf A}_{j_3Q}}^T\left( 
{{\bf A}_{j_2I}} {{\bf A}_{j_2I}}^T + 
{{\bf A}_{j_2Q}} {{\bf A}_{j_2Q}}^T
\right ){{\bf A}_{j_1Q}} &=& {\bf D}^{'}_{j_1,j_2,j_3}, \hspace{8mm}\\
\label{ssddstbc2}
{{\bf A}_{j_1I}}^T\left( 
{{\bf A}_{j_2I}} {{\bf A}_{j_2Q}}^T + 
{{\bf A}_{j_2Q}} {{\bf A}_{j_2I}}^T
\right ){{\bf A}_{j_3Q}} + 
{{\bf A}_{j_3Q}}^T\left( 
{{\bf A}_{j_2I}} {{\bf A}_{j_2Q}}^T + 
{{\bf A}_{j_2Q}} {{\bf A}_{j_2I}}^T
\right ){{\bf A}_{j_1I}} + & &\nonumber \\
{{\bf A}_{j_1Q}}^T\left( 
{{\bf A}_{j_2I}} {{\bf A}_{j_2Q}}^T + 
{{\bf A}_{j_2Q}} {{\bf A}_{j_2I}}^T
\right ){{\bf A}_{j_3I}} + 
{{\bf A}_{j_3I}}^T\left( 
{{\bf A}_{j_2I}} {{\bf A}_{j_2Q}}^T + 
{{\bf A}_{j_2Q}} {{\bf A}_{j_2I}}^T
\right ){{\bf A}_{j_1Q}} &=& {\bf D}^{''}_{j_1,j_2,j_3}, 
\end{eqnarray}
}
where $ {\bf D}^{'}_{j_1,j_2,j_3} $ and  $ {\bf D}^{''}_{j_1,j_2,j_3}$
are block diagonal matrices of the form in (\ref{blockdiagonal}).
\end{thm}
{\em Proof:} First we show the sufficiency part. It can be 
It can be
seen that the matrices
${\bf B}_j^{\prime} = G  \sqrt{E_1} {\bf A}_j \, \left|h_{sj}\right|$,
$j=1,2,\cdots,N$
satisfy the condition (\ref{singlesymx}) in spite of the fact that
$\left|h_{sj} \right|$ are random variables
(since ${\bf B}_j^{\prime}$ matrices are scaled versions of the
${\bf A}_j$ matrices). Let 
${\bf H}_{eq}^{(pc)} = G \sqrt{E_1} \sum_{j=1}^{N} |h_{sj}| h_{jd} {\bf A}_j$. 
It can be seen that 
$\Re\left({{\bf H}^{(pc)}_{eq}}^{\mathcal{H}} {\bf H}^{(pc)}_{eq} \right)$ 
is block diagonal of the form in (\ref{blockdiagonal}). 
This implies that
each element of the $K\times 1$ vector
$\Re \left( {{\bf H}^{(pc)}_{eq}}^{\mathcal{H}} {\bf y} \right)$
is affected by only one information symbol (i.e., there will be no
information symbol entanglement in each element).
Hence, for SSD, it suffices
to show that noise in each of these terms are uncorrelated, i.e.,
the DSTBC is SSD {\em iff} 
{\small
$E \left [ \Re \left ( {{\bf H}_{eq}^{(pc)}}^{\mathcal{H}} \tilde{\bf z}_d \right ) 
\Re \left ( {{\bf H}_{eq}^{(pc)}}^{\mathcal{H}} \tilde{\bf z}_d \right )^T \right ]$} 
is a block diagonal matrix of the form (\ref{blockdiagonal}). 
Expanding
{\small
$E \left [ \Re \left ( {{\bf H}_{eq}^{(pc)}}^{\mathcal{H}} \tilde{\bf z}_d \right ) \Re \left ( {{\bf H}_{eq}^{(pc)}}^{\mathcal{H}} \tilde{\bf z}_d \right )^T \right ] $}, we arrive, after some manipulation, at
{\footnotesize
\begin{eqnarray}
\label{longeqn}
E \left [  \Re \left ( {\bf H}_{eq}^{\mathcal{H}} \tilde{\bf z}_d \right )
\Re \left ( {\bf H}_{eq}^{\mathcal{H}} \tilde{\bf z}_d \right )^T \right ] &=&
\sum_{j_1 =1}^{N}\sum_{j_2 =1}^{N}\sum_{j_3 =1}^{N} |h_{sj_1}||h_{j_2d}|^2 |h_{sj_3}| \left ( h_{j_1dI} h_{j_3dI} + h_{j_1dQ} h_{j_3dQ} \right ) \nonumber \\
& &
\hspace{-20mm}
\Bigg [
{{\bf A}_{j_1I}}^T\left(
{{\bf A}_{j_2I}} {{\bf A}_{j_2I}}^T +
{{\bf A}_{j_2Q}} {{\bf A}_{j_2Q}}^T
\right ){{\bf A}_{j_3I}} 
+ \,\, {{\bf A}_{j_3I}}^T\left(
{{\bf A}_{j_2I}} {{\bf A}_{j_2I}}^T +
{{\bf A}_{j_2Q}} {{\bf A}_{j_2Q}}^T
\right ){{\bf A}_{j_1I}}   \nonumber \\
& &
\hspace{-20mm}
+ \,\, {{\bf A}_{j_1Q}}^T\left(
{{\bf A}_{j_2I}} {{\bf A}_{j_2I}}^T +
{{\bf A}_{j_2Q}} {{\bf A}_{j_2Q}}^T
\right ){{\bf A}_{j_3Q}} +
{{\bf A}_{j_3Q}}^T\left(
{{\bf A}_{j_2I}} {{\bf A}_{j_2I}}^T +
{{\bf A}_{j_2Q}} {{\bf A}_{j_2Q}}^T
\right ){{\bf A}_{j_1Q}}
\Bigg] \nonumber \\
& &
\hspace{-20mm}
+ \,\,
\sum_{j_1 =1}^{N}\sum_{j_2 =1}^{N}\sum_{j_3 =1}^{N} |h_{sj_1}||h_{j_2d}|^2 |h_{sj_3}| \left ( h_{j_1dI} h_{j_3dQ} + h_{j_1dQ} h_{j_3dI} \right ) \nonumber \\
& &
\hspace{-20mm}
\Bigg [
{{\bf A}_{j_1I}}^T\left(
{{\bf A}_{j_2I}} {{\bf A}_{j_2Q}}^T +
{{\bf A}_{j_2Q}} {{\bf A}_{j_2I}}^T
\right ){{\bf A}_{j_3Q}}
+ \,\,
{{\bf A}_{j_3Q}}^T\left(
{{\bf A}_{j_2I}} {{\bf A}_{j_2Q}}^T +
{{\bf A}_{j_2Q}} {{\bf A}_{j_2I}}^T
\right ){{\bf A}_{j_1I}}   \nonumber \\
& &
\hspace{-20mm}
+ \,\,
{{\bf A}_{j_1Q}}^T\left(
{{\bf A}_{j_2I}} {{\bf A}_{j_2Q}}^T +
{{\bf A}_{j_2Q}} {{\bf A}_{j_2I}}^T
\right ){{\bf A}_{j_3I}} +
{{\bf A}_{j_3I}}^T\left(
{{\bf A}_{j_2I}} {{\bf A}_{j_2Q}}^T +
{{\bf A}_{j_2Q}} {{\bf A}_{j_2I}}^T
\right ){{\bf A}_{j_1Q}}
\Bigg ] \nonumber \\
& = &
\sum_{j_1 =1}^{N}\sum_{j_2 =1}^{N}\sum_{j_3 =1}^{N} |h_{sj_1}||h_{j_2d}|^2 |h_{sj_3}| \left ( h_{j_1dI} h_{j_3dI} + h_{j_1dQ} h_{j_3dQ} \right )
{\bf D}^{'}_{j_1,j_2,j_3} \nonumber \\
& &
+\,\,
\sum_{j_1 =1}^{N}\sum_{j_2 =1}^{N}\sum_{j_3 =1}^{N} |h_{sj_1}||h_{j_2d}|^2 |h_{sj_3}| \left ( h_{j_1dI} h_{j_3dQ} + h_{j_1dQ} h_{j_3dI} \right )
{\bf D}^{''}_{j_1,j_2,j_3},
\end{eqnarray}
}
where, in terms of notation, $h_{j_1dI}$ and $h_{j_1dQ}$ denote the
real and imaginary parts of the channel gains from the relay $j_1$ to
destination $d$ (i.e., the real and imaginary parts of $h_{j_1d}$),
respectively. Since (\ref{longeqn}) turns out to be a linear combination
of the ${\bf D}^{'}_{j_1,j_2,j_3}$ and ${\bf D}^{''}_{j_1,j_2,j_3}$
matrices in (\ref{ssddstbc1}) and  (\ref{ssddstbc2}),
the covariance matrix
is of the form (\ref{blockdiagonal}). Hence, along with (\ref{singlesymx})
the conditions in  (\ref{ssddstbc1}) and (\ref{ssddstbc2}) constitute a set
of sufficient conditions.

To show the ``necessary part,'' since the terms 
$h_{sj_1}||h_{rj_2}|^2 |h_{sj_3}| 
(h_{rj_1I} h_{rj_3I} + h_{rj_1Q} h_{rj_3Q} ) $ and 
$h_{sj_1}||h_{rj_2}|^2 |h_{sj_3}| 
(h_{rj_1I} h_{rj_3Q} + h_{rj_1Q} h_{rj_3I} ) $ 
are independent and if the co-variance matrix has to be block diagonal for 
all the realizations of $h_{sj}$ and $ h_{rj} $, then the conditions  in 
(\ref{ssddstbc1}) and (\ref{ssddstbc2}) have to be necessarily satisfied. 
Also, in the similar lines of the proof for {\em Theorem 1}, it can be 
deduced that ${\bf B}_j^{\prime}$ satisfying condition (\ref{singlesymx}) 
is necessary 
to achieve un-entanging of information symbols in the elements
of the vector
$ \Re \left ( {{\bf H}^{(pc)}_{eq}}^{\mathcal{H}} {\bf y} \right )$.
$\square$ 

In \cite{RaR3}, partially-coherent distributed set up has been studied 
and a sufficient condition has been identified for a distributed STBC 
to be SSD. In the following corollary, it is shown that Theorem \ref{thm2} 
subsumes this sufficient condition as a special case.

\vspace{-2mm}
\begin{cor} The sufficient condition in \cite{RaR3}, i.e., the noise
co-variance to be a scaled identity matrix, is a subset of the conditions 
(\ref{ssddstbc1}) and (\ref{ssddstbc2}).
\end{cor}
{\em Proof: } It can be observed that the ${\bf Z}_j $ matrix in \cite{RaR3}, 
when written in our notation, is  $ {\bf Z}_j = 
\left [ \begin{array} {c} {\bf A}_{jI} \\ {\bf A}_{jQ} \end{array} \right ]$. 
Hence, if $ {\bf Z}_j {\bf Z}_j^T = \alpha {\bf I}\,\, \forall j $, 
where $\alpha$ is a scalar, then,  
${\bf A}_{jI}  {{\bf A}_{jI}}^T = \alpha{\bf I} $,
${\bf A}_{jQ}  {{\bf A}_{jQ}}^T = \alpha{\bf I} $,
${\bf A}_{jQ}  {{\bf A}_{jI}}^T = {\bf 0},$ and
${\bf A}_{jI}  {{\bf A}_{jQ}}^T = {\bf 0} $.
Substituting this in (\ref{ssddstbc1}) and (\ref{ssddstbc2}), we get the left 
hand side of (\ref{ssddstbc1}) to be
\begin{eqnarray*}
\alpha \left( {{\bf A}_{j_1I}}^T {{\bf A}_{j_3I}} + 
{{\bf A}_{j_3I}}^T {{\bf A}_{j_1I}} +
{{\bf A}_{j_1Q}}^T {{\bf A}_{j_3Q}} + 
{{\bf A}_{j_3Q}}^T {{\bf A}_{j_3Q}} \right),
\end{eqnarray*}
which, by (\ref{singlesymx}), is always a block diagonal matrix of the form
(\ref{blockdiagonal}). Also, the left hand side of (\ref{ssddstbc2}) is 
{\bf 0}.
Hence, ${\bf A}_{jI}  {{\bf A}_{jI}}^T = 
{\bf A}_{jQ}  {{\bf A}_{jQ}}^T = \alpha{\bf I} $ and
$ {\bf A}_{jI}  {{\bf A}_{jQ}}^T = 
{\bf A}_{jQ}  {{\bf A}_{jI}}^T = {\bf 0} \,\,
 \forall j $ is a sufficient
condition for a DSTBC to be SSD.  $\square$ 

In \cite{RaR3}, it is shown that the 8-antenna code given by (\ref{cliff8}), 
which we
denote by $RR_8,$ does not satisfy the sufficient condition discussed in 
that paper for SSD in PCRC, and hence not claimed to be SSD. However, it 
can be verified that $RR_8$ satisfies (\ref{singlesymx}), 
(\ref{ssddstbc1}) and (\ref{ssddstbc2}), and hence SSD-DSTBC-PCRC.

\vspace{-2mm}
\begin{scriptsize}
\begin{eqnarray}
\label{cliff8}
\mbox{\hspace{-10mm}}
RR_8 & \mbox{\hspace{-3mm}} = &  \mbox{\hspace{-3mm}}
\left ( \begin{array} {cccccccc}
 x_{1I} - {\bf j} x_{4Q} &  x_{2I} + {\bf j} x_{3I} &
 x_{4I} + {\bf j} x_{1Q} & -x_{3Q} + {\bf j} x_{2Q} & 0&0&0&0\\
-x_{2I} - {\bf j} x_{3I} &  x_{1I} + {\bf j} x_{4Q} &
-x_{3Q} - {\bf j} x_{2Q} & -x_{4I} + {\bf j} x_{1Q} & 0&0&0&0 \\
-x_{4I} - {\bf j} x_{1Q} &  x_{3Q} - {\bf j} x_{2Q} &
 x_{1I} - {\bf j} x_{4Q} &  x_{2I} + {\bf j} x_{3I} & 0&0&0&0 \\
 x_{3Q} + {\bf j} x_{2Q} &  x_{4Q} - {\bf j} x_{1Q} &
-x_{2I} - {\bf j} x_{3I} &  x_{1I} + {\bf j} x_{4Q} & 0&0&0&0\\
0 & 0 & 0 & 0 &
 x_{1I} - {\bf j} x_{4Q} &  x_{2I} + {\bf j} x_{3I} &
 x_{4I} + {\bf j} x_{1Q} & -x_{3Q} + {\bf j} x_{2Q} \\
0 & 0 & 0 & 0 &
-x_{2I} - {\bf j} x_{3I} &  x_{1I} + {\bf j} x_{4Q} &
-x_{3Q} - {\bf j} x_{2Q} & -x_{4I} + {\bf j} x_{1Q} \\
0 & 0 & 0 & 0 &
-x_{4I} - {\bf j} x_{1Q} &  x_{3Q} - {\bf j} x_{2Q} &
 x_{1I} - {\bf j} x_{4Q} &  x_{2I} + {\bf j} x_{3I} \\
0 & 0 & 0 & 0 &
 x_{3Q} + {\bf j} x_{2Q} &  x_{4Q} - {\bf j} x_{1Q} &
-x_{2I} - {\bf j} x_{3I} &  x_{1I} + {\bf j} x_{4Q}
\end{array} \right). \mbox{\hspace{6mm}}
\end{eqnarray}
\end{scriptsize}

\vspace{-4mm}
\subsection{Invariance of SSD under coordinate interleaving}
\vspace{-4mm}
In this subsection, we show that the property of SSD of a DSTBC for PCRC is 
invariant under coordinate interleaving of the data symbols. To illustrate 
the usefulness of this result we first show the following lemma.
\vspace{-2mm}
\begin{lem} 
\label{lem2}
If ${\bf G}(x_1,\cdots,x_{T_1}) $ is a SSD design
in $ T_1$ variables and $ N $ transmit nodes that satisfies
(\ref{singlesymx}), (\ref{ssddstbc1}) and (\ref{ssddstbc2}),
then the design in $ 2T_1$ variables and $ 2N $  transmit nodes 
given by
\begin{eqnarray}
\label{gbar}
\bar{\bf G}(x_1,\cdots,x_{2T_1}) = \left [
\begin{array} {cc}
{\bf G}(x_1,\cdots,x_{T_1}) & {\bf 0} \\
{\bf 0} & {\bf G}(x_{T_1 + 1},\cdots,x_{2T_1})
\end{array}
\right ]
\end{eqnarray}
also satisfies  (\ref{singlesymx}), (\ref{ssddstbc1}) and (\ref{ssddstbc2}).
\end{lem}
{\em Proof: } If $ {\bf A}_j, \,\, 1 \leq j \leq N $ are the relay matrices 
of  ${\bf G}$, then the corresponding  $ \bar{\bf A}_j $
matrices for $ \bar{{\bf G}}$ are
$ \bar{\bf A}_j =
\left [ \begin{array} {cc} {\bf A}_j & {\bf 0} \\ {\bf 0} & {\bf 0}  
\end{array} \right ]
,\,\, 1 \leq j \leq N $ and
$ \bar{\bf A}_j =
\left [ \begin{array} {cc}  {\bf 0} & {\bf 0} \\ {\bf 0} &  {\bf A}_j 
\end{array} \right ]
,\,\, N + 1 \leq j \leq 2N $. It is easily verified that if $ {\bf A}_j $
satisfies  (\ref{singlesymx}), (\ref{ssddstbc1}) and (\ref{ssddstbc2}),
then so do the matrices $  \bar{\bf A}_j. $  $\square$  

As an example, if we choose ${\bf G}(x_1,x_2)$ to be the Alamouti code in 
the lemma above then we get the code 
\begin{eqnarray}
\label{doubleala}
 \left [
\begin{array} {rrrr}
 x_1 & x_2 &  0 & 0 \\
-x_2^* & x_1^* & 0 & 0 \\
0 & 0 & x_3 & x_4 \\
0 & 0 & -x_4^* & x_3^*
\end{array}
\right ].
\end{eqnarray}
This code is SSD for PCRC. Note that a 4-antenna COD has only rate only 
$\frac{3}{4}$ whereas this code has rate 1.  However, it is easily shown  
that this code does not give full-diversity. But, coordinate interleaving 
for this example results in $CIOD_4$ which gives full-diversity for any 
signal set with coordinate product distance zero, and we have already 
seen that $CIOD_4$ has the SSD property for PCRC.

The following theorem shows that it is the property of coordinate interleaving to leave the SSD property of any arbitrary STBC for PCRC intact.
\vspace{-2mm}
\begin{thm} 
\label{thm4}
If an STBC with $K$ variables $x_1,x_2,\cdots,x_K,$ satisfy  
(\ref{singlesymx}), (\ref{ssddstbc1}) and (\ref{ssddstbc2}), the SSD property
is unaffected by doing arbitrary coordinate interleaving among all real
and imaginary components of $x_i$.\footnote{It should be noted that neither 
the source nor the relay does an
explicit interleaving, but the net effect of the relay matrices is such that
the output of relays is an interleaved version of the information symbols.}
\end{thm}
\vspace{-2mm}
{\em Proof:} The data-symbol vector in (\ref{datavector}) after interleaving
can be written as
\begin{eqnarray*}
\tilde{\bf x} & = & \tilde{\bf I} \, {\bf x}
\end{eqnarray*}
where $\tilde{\bf I}$ is the interleaving matrix which is a permutation 
matrix obtained by permuting the rows (/columns) of the identity matrix 
{\bf I} to reflect the coordinate interleaving operation. It can be easily 
checked that $ \tilde{\bf I}^2={\bf I}.$ Also, if $ {\bf D} $ is a block 
diagonal matrix of the form (\ref{2bdiagonal}), then so is the matrix  
$ \tilde{\bf I} {\bf D}  \tilde{\bf I}.$ Hence, for PCRC with co-ordinate 
interleaving (\ref{stacky}) can be written as 
\begin{eqnarray}
\label{stackxnew}
{\bf c}_{j} & = & {\bf A}_j \widehat{\bf v}_{j} \nonumber \\
& = & \underbrace{G \, \sqrt{E_1} {\bf A}_j |h_{sj}|\tilde{\bf I}}_{{\bf B}_j^{'}}\, {\bf x} \, + \,{\bf A}_j \, {\bf \widehat{z}}_{j}, 
\end{eqnarray} 
which means that after interleaving, the equivalent linear processing
matrix is $ {\bf A}_j  \tilde{\bf I} $. It is easily verified that
if $ {\bf A}_j  $ satisfies (\ref{singlesymx}), 
(\ref{ssddstbc1}) and (\ref{ssddstbc2}), then so does 
${\bf A}_j  \tilde{\bf I}$ also.
$\square$ 

As an example, consider the Alamouti code 
$\left[\begin{array}{rr}
x_1 & x_2 \\
-x_2^* & x_1^*
\end{array}\right],$
whose relay matrices are given by (\ref{AA}). For this case,
$N=T_1=T_2=2.$ The permutation matrix $\tilde{\bf I}$ for the 
coordinate interleaving operation is 
$\left[\begin{array}{cccc} 
1 & 0 & 0 & 0 \\
0 & 0 & 0 & 1 \\
0 & 0 & 1 & 0 \\
0 & 1 & 0 & 0 
\end{array}\right].$ 
The relay matrices for the coordinate interleaved code are
\begin{eqnarray}
\label{AAnew}
{\bf A}_1 \tilde{\bf I}= \left [ \begin{array}{cccc} 1 & 0 & 0 & {\bf j} \\
0 & {\bf j} & -1 & 0 \end{array} \right ]
 & \mbox{ and } &
{\bf A}_2 \tilde{\bf I}= \left [ \begin{array}{cccc} 0 &  {\bf j} & 1 & 0\\
1 & 0& 0 & -{\bf j}  \end{array} \right ],
\end{eqnarray}
and the resulting code is 
$\left[\begin{array}{rr}
x_{1I}+jx_{2Q} & x_{2I}+jx_{1Q} \\
-x_{2I}+jx_{1Q} & x_{1I}-jx_{2Q} 
 \end{array}\right]= 
\left[\begin{array}{rr}
\tilde{x}_1 & \tilde{x}_2 \\
-\tilde{x}_2^* & \tilde{x}_1^*
 \end{array}\right].$
Also, for the code in (\ref{doubleala}) which is SSD for PCRC, if we choose
the permutation matrix $\tilde{\bf I}$ as
\begin{eqnarray}
\tilde{\bf I} = \left [ \begin{array} {cccccccc}
1 & 0 & 0 & 0 & 0 & 0 & 0 & 0 \\
0 & 0 & 0 & 0 & 0 & 1 & 0 & 0 \\
0 & 0 & 1 & 0 & 0 & 0 & 0 & 0 \\
0 & 0 & 0 & 0 & 0 & 0 & 0 & 1 \\
0 & 0 & 0 & 0 & 1 & 0 & 0 & 0 \\
0 & 1 & 0 & 0 & 0 & 0 & 0 & 0 \\
0 & 0 & 0 & 0 & 0 & 0 & 1 & 0 \\
0 & 0 & 0 & 1 & 0 & 0 & 0 & 0 \\
\end{array} \right ],
\end{eqnarray}
the resulting code is  given by 
\begin{eqnarray}
\left[\begin{array}{rrrr}
x_{1I}+jx_{3Q} & x_{2I}+jx_{4Q} & 0 & 0\\
-x_{2I}+jx_{4Q} & x_{1I}-jx_{3Q} & 0 & 0 \\ 
0 & 0 & x_{3I}+jx_{1Q} & x_{4I}+jx_{2Q} \\ 
0 & 0 & -x_{4I}+jx_{2Q} & x_{3I}-jx_{1Q} \\ 
 \end{array}\right]= 
\left[\begin{array}{rrrr}
\tilde{x}_1 & \tilde{x}_2  & 0 & 0\\
-\tilde{x}_2^* & \tilde{x}_1^* & 0 & 0 \\
0 & 0 & \tilde{x}_3^* & \tilde{x}_4^* \\
0 & 0 & -\tilde{x}_4^* & \tilde{x}_3^* \\
 \end{array}\right],
\end{eqnarray}
which is $CIOD_4$. Hence, $CIOD_4 $ is also SSD for PCRC. In general, if we
have a code with $K$ complex information symbols which is SSD for PCRC,
then we can generate $(2K)!$ codes which are SSD for PCRC by coordinate
interleaving.
\vspace{-4mm}
\subsection{A class of rate-$\frac{1}{2}$ SSD DSTBCs}
\vspace{-4mm}
All the classes of codes discussed so far are STBCs from square designs. 
It is well known that the rate of square SSD codes for co-located MIMO
systems falls exponentially as 
the number of antennas increases. In this subsection, it is shown that if 
non-square designs are used then SSD codes for PCRCs can be achieved with 
rate $\frac{1}{2}$ for any number of antennas.

It is well known \cite{TJC} that real orthogonal designs (RODs) with rate 
one exist for any number of antennas and these are non-square designs for 
more than 2 antennas and the delay increases exponentially with the number 
of antennas. Using these RODs, in \cite{TJC}, a class of rate $\frac{1}{2}$ 
complex orthogonal designs for any number of antennas is obtained as follows: 
If ${\bf G}$ is a $p \times N$ rate one ROD, where $p$ denotes the delay and 
$N$ denotes the number of antennas with variables $x_1,x_2, \cdots, x_p,$ 
then, denoting by ${\bf G}^*$ the complex design obtained by replacing 
$x_i$ with $x_i^*, ~i=1,2,\cdots,p,$ the design
$\left[
\begin{array} {c} 
{\bf G} \\ {\bf G}^*
\end{array} 
\right]$
is a $2p \times N$ rate-$\frac{1}{2}$ COD. We refer to this construction 
as stacking construction. The following theorem asserts that the rate 
$\frac{1}{2}$ CODs by stacking construction are SSD for PCRC. 
\vspace{-2mm}
\begin{thm} 
\label{ratehalfthm} 
The rate-1/2 CODs, constructed from rate one RODs by stacking construction  
\cite{TJC} are SSD-DSTBC-PCRC.
\end{thm}
\vspace{-2mm}
{\em Proof: } Let ${\bf G}_c$ be the rate-1/2 COD obtained from a 
$p \times N$  ROD  ${\bf G}$ by stacking construction, i.e.,  
\begin{eqnarray}
{\bf G}_c = \left [  \begin{array} {c} {\bf G} \\ {\bf G}^* 
\end{array} \right ].
\end{eqnarray}
Let the $ p \times p $ real matrices $\hat{{\bf A}}_j \, j=1,\cdots,N $ 
generate the columns of ${\bf G}$, i.e.,
\begin{eqnarray}
{\bf G}& = & \left [ \hat{{\bf A}}_1 {\bf x}, \hat{{\bf A}}_2 {\bf x}, \cdots,
\hat{{\bf A}}_N {\bf x}  \right ],
\end{eqnarray}
where ${\bf x}$ is the $p \times 1$ real data vector and the matrices
$ \hat{{\bf A}}_j $ denote the column vector representation matrices used 
in \cite{Lia}.
By the definition of RODs, ${\bf G}^T {\bf G}=\left ({\bf x}^T{\bf x} \right){\bf I}$. 
This implies that 
\begin{eqnarray}
\hat{\bf A}_j^T \hat{\bf A}_j &=& {\bf I}, \,\, j = 1,\cdots,N \nonumber \\
\hat{\bf A}_j^T \hat{\bf A}_i &=& - \hat{\bf A}_i^T \hat{\bf A}_j, 
\,\, i,j = 1,\cdots,N, i \neq j.
\label{rod_condn}
\end{eqnarray}
It is noted that the Hurwitz-Radon family of matrices satisfy 
(\ref{rod_condn}) and explicit construction for any $N$ is given in 
\cite{TJC}. It is noted that the representation in \cite{TJC} is 
different from the column vector representation used in this paper. An 
important consequence is that the Hurwitz-Radon family of matrices 
satisfy the conditions
\begin{eqnarray}
\hat{\bf A}_j^T \hat{\bf A}_j &=& {\bf I}, \,\, j = 1,\cdots,N \nonumber \\
\hat{\bf A}_j^T  &=&  -\hat{\bf A}_j, \,\, j = 1,\cdots,N \nonumber \\
\hat{\bf A}_j \hat{\bf A}_i &=& - \hat{\bf A}_i \hat{\bf A}_j, 
\,\, i,j = 1,\cdots,N, i \neq j,
\end{eqnarray}
and hence $ \hat{\bf A}_j \hat{\bf A}_j^T = {\bf I} \,\, \forall j,$ which we 
will use in our proof. Viewing ${\bf G}_c$ as a $T_2 \times N$ distributed 
STBC with $T_1=p$ and $T_2=2p$, the $T_2 \times 2T_1$ relay matrices 
$ {\bf A}_j $ of ${\bf G}_c $ have the structure
\begin{eqnarray}
{\bf A}_{jI} = 
\left (  \begin{array} {c} {\bf U}_j \\ {\bf U}_j \end{array} \right ) \,\,\,
\mbox{and} \,\,\,
{\bf A}_{jQ} = 
\left (  \begin{array} {r} {\bf V}_j \\ {\bf -V}_j \end{array} \right ).
\end{eqnarray}
Since ${\bf G}_c$ is constructed from a ROD, the coefficients of real and 
imaginary components are same, i.e., the matrices ${\bf U}_j$ and ${\bf V}_j$ 
have the form
\begin{equation}
{\bf U}_j = \left[{\bf\gamma}_{1,j},{\bf 0},{\bf\gamma}_{2,j}, {\bf 0},
\cdots,{\bf\gamma}_{T_1,j}, {\bf 0 }\right ],\,\,\,\,\,\,\,\,\,\,\,
{\bf V}_j = \left[{{\bf 0}, \bf\gamma}_{1,j},{\bf 0}, 
{\bf\gamma}_{2,j},\cdots,{\bf 0},{\bf\gamma}_{T_1,j}\right ],
\end{equation} 
with $ \gamma_{i,j}$ are column vectors of $\hat{\bf A}_j$. Since 
$\hat{\bf A}_j \hat{\bf A}_j^T = {\bf I} \,\, \forall j$, it is easily
verified that $ {\bf U}_j {\bf U}_j^T =  {\bf I}$ and 
$ {\bf V}_j {\bf V}_j^T =  {\bf I} \,\, \forall j $. It is also easily seen 
that  $ {\bf U}_j {\bf V}^T_j = {\bf 0} $ and $ {\bf V}_j {\bf U}^T_j = 
{\bf 0} $. Hence, we have
\begin{eqnarray}
{\bf A}_{jI} {{\bf A}_{jI}}^T + {\bf A}_{jQ} {{\bf A}_{jQ}}^T 
& = & 2  {\bf I} \nonumber \\
{\bf A}_{jI} {{\bf A}_{jQ}}^T + {\bf A}_{jQ} {{\bf A}_{jI}}^T 
& = & {\bf 0} 
\end{eqnarray}
Substituting this in (\ref{ssddstbc1}), we get the left hand side of 
(\ref{ssddstbc1}) to be
\begin{eqnarray}
2 \left( {{\bf A}_{j_1I}}^T {{\bf A}_{j_3I}} + 
{{\bf A}_{j_3}}^T {{\bf A}_{j_1I}} +
{{\bf A}_{j_1Q}}^T {{\bf A}_{j_3Q}} + 
{{\bf A}_{j_3Q}}^T {{\bf A}_{j_3Q}} \right),
\end{eqnarray}
which, by (\ref{singlesymx}), is always a block diagonal matrix of the form
(\ref{blockdiagonal}). Also the left hand side of (\ref{ssddstbc2}) is 
${\bf 0}$. Hence, ${\bf G}_c$ is SSD for PCRC. $\square$ \\

\vspace{-6mm}
In \cite{YiK}, it is shown that if the $N$ relays do not have any CSI 
and the destination has all the CSI, then an upper bound on the rate 
of distributed SSD codes is $\frac{2}{N},$ which decreases rapidly as 
the number of relays increases. However, Theorem \ref{ratehalfthm} shows 
that, if the relay knows only the phase information of the source-relay 
channels then the lower bound on the rate of the distributed SSD codes is 
$\frac{1}{2}$ which is independent of the number of relays. For example, 
the ROD part of such rate-1/2 SSD DSTBCs for PCRC for 10 and 12 
relays are given in (\ref{rod10}) and (\ref{rod12}), respectively, where 
Hurwitz-Radon construction yields the $32\times 10$ matrix in (\ref{rod10})
for 10 relays and the $64\times 12$ matrix in (\ref{rod12}) for 12 relays.
\vspace{-4mm}
\subsection{Full-diversity, single-symbol non-ML detection }
\vspace{-4mm}
\begin{thm} 
\label{nonMLSSD}The PCRC system given by
(\ref{rx2}) achieves full diversity irrespective of whether the total noise
($\tilde{{\bf z}}_d  $) is correlated or not, if the STBC achieves full 
diversity in the co-located case and condition (\ref{singlesymx}) is 
satisfied.
\end{thm}
\vspace{-4mm}
{\em Proof: }
Since the noise $\tilde{{\bf z}}_d $ is not assumed to be uncorrelated, 
the optimal detection of ${\bf x}$ in the maximum likelihood sense is given by 
\begin{eqnarray}
\label{ml_opt}
\widehat{\bf x} &=& \mbox{arg min} \,\, 
( {\bf y} - {\bf H}^{(pc)}_{eq}{\bf x} )^{\mathcal{H}} {\bf \Omega}^{-1} 
( {\bf y} - {\bf H}^{(pc)}_{eq} {\bf x} ),
\end{eqnarray}
where ${\bf \Omega}$ is co-variance matrix of the noise, given by 
${\bf \Omega} = E\{ \tilde{{\bf z}}_d  \tilde{{\bf z}}_d  ^{\mathcal{H}}  \} $. We consider
the sub-optimal metric (ignoring ${\bf \Omega}^{-1}$) 
\begin{eqnarray}
\label{ml_sub_opt}
\widehat{\bf x} &=& \mbox{arg min} \,\, 
({\bf y} - {\bf H}^{(pc)}_{eq}{\bf x})^{\mathcal{H}} 
({\bf y} - {\bf H}^{(pc)}_{eq}{\bf x}),
\end{eqnarray}
and show that this decision metric achieves full diversity. Proceeding on
the similar lines for the proof for the co-located case, the pair-wise
error probability is upper bounded by
\begin{eqnarray}
\label{d_cb1}
P\left( {\bf x}_1 \rightarrow {\bf x}_2 \right ) \leq 
E \left\{ e^{-d^2({\bf x}_1,{\bf x}_2) E_t/4 } \right\},
\end{eqnarray}
where the Euclidean distance in (\ref{d_cb1}) can be 
written as
\begin{eqnarray}
\label{d_cb2}
d^2({\bf x}_1,{\bf x}_2) = 
({\bf x}_2 - {\bf x}_1)^T \Re \left ( 
{{\bf H}^{(pc)}_{eq}}^{\mathcal{H}} {\bf H}^{(pc)}_{eq} \right )({\bf x}_2 - {\bf x}_1).
\end{eqnarray}
Since (\ref{singlesymx}) is satisfied, this can be written as sum of  
$T_1$ terms as
\begin{eqnarray}
d^2({\bf x}_1,{\bf x}_2) &= &
 \sum_{i=1}^{T_1} \Delta {{\bf x}^{(i)}}^T \left ( 
\sum_{j=1}^{N}  |h_{sj}|^2  |h_{jd}|^2 {\bf D}^{(1)}_{j,i}
\right ) \Delta {\bf x}^{(i)} \\
& = &
 \sum_{j=1}^{N} |h_{sj}|^2 |h_{jd}|^2  \left (
\sum_{i=1}^{T_1} \Delta {{\bf x}^{(i)}}^T  
{\bf D}^{(1)}_{j,i}  \Delta {\bf x}^{(i)}
\right ). 
\label{eq_here2}
\end{eqnarray}
Substituting (\ref{eq_here2}) in (\ref{d_cb1}) and evaluating the expectation 
with respect to $ |h_{jd}|^2 $, we get
\begin{eqnarray}
\label{d_cb4}
P\left( {\bf x}_1 \rightarrow {\bf x}_2 \right | h_{sj} ) \leq 
\prod_{j=1}^N\left ( \frac{1}{1 + 
|h_{sj}|^2   \sum_{i=1}^{T_1} \Delta {{\bf x}^{(i)}}^T  
{\bf D}^{(1)}_{j,i}  \Delta {\bf x}^{(i)} E_t/4,
} \right ),
\end{eqnarray}
which, for high SNRs, could be approximated as
\begin{eqnarray}
P\left( {\bf x}_1 \rightarrow {\bf x}_2 | h_{sj}  \right ) \leq 
\prod_{j=1}^N\left ( \frac{1}{ \sum_{i=1}^{T_1} \Delta {{\bf x}^{(i)}}^T  
{\bf D}^{(1)}_{j,i}  \Delta {\bf x}^{(i)} E_t/4
} \right )
\prod_{j=1}^N\left ( \frac{1}{|h_{sj}|^2 } \right ).
\end{eqnarray}
Now, evaluating the expectation with respect to $ |h_{sj}|, $ we get
\begin{eqnarray}
\label{d_cb5}
P\left( {\bf x}_1 \rightarrow {\bf x}_2 \right ) \leq 
\prod_{j=1}^N\left ( \frac{1}{ \sum_{i=1}^{T_1} \Delta {{\bf x}^{(i)}}^T  
{\bf D}^{(1)}_{j,i}  \Delta {\bf x}^{(i)} E_t/4
} \right )
\left ( {\bf Ei}(0) \right )^N,
\end{eqnarray}
where $ {\bf Ei}(x) $  is the exponential integral
$ \int_x^{\infty} \frac{e^{-t}}{t} dt $. From (\ref{d_cb5}), it is clear that
the condition for achieving maximum diversity is identical to that of 
co-located MIMO (\ref{cb4}).  $\square$ 

Theorem \ref{nonMLSSD} means that by using any STBC 
which satisfies the conditions (\ref{singlesymx}) and achieves full 
diversity in co-located MIMO system, it is possible to do decoding of 
one symbol at a time and achieve full diversity, though not optimal in 
the ML sense, in a distributed setup with phase compensation done at 
the relay, even if {\it (\ref{ssddstbc1}) and (\ref{ssddstbc2}) are not 
satisfied }. For example, the $CIOD_8$ is SSD and gives full-diversity 
in a co-located 8-transmit antenna system for any signal set with 
coordinate product distance (CPD) not equal to zero, and is not SSD 
for PCRC since it does not satisfy the (\ref{ssddstbc1}) and 
(\ref{ssddstbc2}). However, according to Theorem \ref{nonMLSSD} a SSD 
decoder for $CIOD_8$ in a PCRC will result in full-diversity of order 8. 

\vspace{-6mm}
\section{Discussion and Simulation Results}
\label{sec5}
\vspace{-6mm}
The results of our necessary and sufficient conditions (\ref{singlesymx}),
(\ref{ssddstbc1}) and (\ref{ssddstbc2}) as well as the sufficient condition 
in \cite{RaR3}, evaluated for various classes of codes for PCRC are shown 
in Table \ref{tab1}. As can be seen from the last column of Table \ref{tab1}, 
the sufficient condition in \cite{RaR3} identifies only $COD_2$ (Alamouti) 
and $CUW_4$ as SSDs for PCRC. However, our conditions (\ref{singlesymx},
(\ref{ssddstbc1}) and (\ref{ssddstbc2}) identify $CIOD_4$, $RR_8$, and 
$COD$s from $ROD$s, in addition to $COD_2$ and $CUW_4$, as SSDs for PCRC 
(4th column of Table \ref{tab1}). It is noted that, $CIOD_4$ being a 
construction by using ${\bf G}=COD_2$ in (\ref{gbar}) and coordinate 
interleaving, it is SSD for PCRC from {\em Lemma \ref{lem2}} and {\em 
Theorem \ref{thm4}}. Similarly, since $RR_8$ code is constructed by using 
${\bf G}=CUW_4$ in (\ref{gbar}), it follows from {\em Lemma \ref{lem2}} 
that $RR_8$ is also SSD for PCRC. Also, CODs from RODs are SSD for PCRC 
from {\em Theorem \ref{ratehalfthm}}. Since $COD_4$, $COD_8$, and $CIOD_8$ 
do not satisfy our conditions, they are not SSD for PCRC.

Next, we present the bit error rate (BER) performance of various classes
of codes without and with phase compensation at the relays (i.e., PCRC). 
For the purposes of the simulation results and discussions in this section, 
we classify the decoding of codes for PCRC into two categories: $i)$ codes 
for which single symbol decoding is ML-optimal; we refer to this decoding 
as ML-SSD; we consider ML-SSD of $COD_2$ and $CIOD_4$, and $ii)$ codes 
which when decoded using single symbol decoding are not ML-optimal, but 
achieve full diversity; we refer to this decoding as non-ML-SSD; 
we consider non-ML-SSD of $COD_4$, $COD_8$, and $CIOD_8$. When no phase 
compensation is done at the relays, we consider ML decoding.

\begin{figure}
\begin{center}
\epsfxsize=10.0cm
\epsfbox{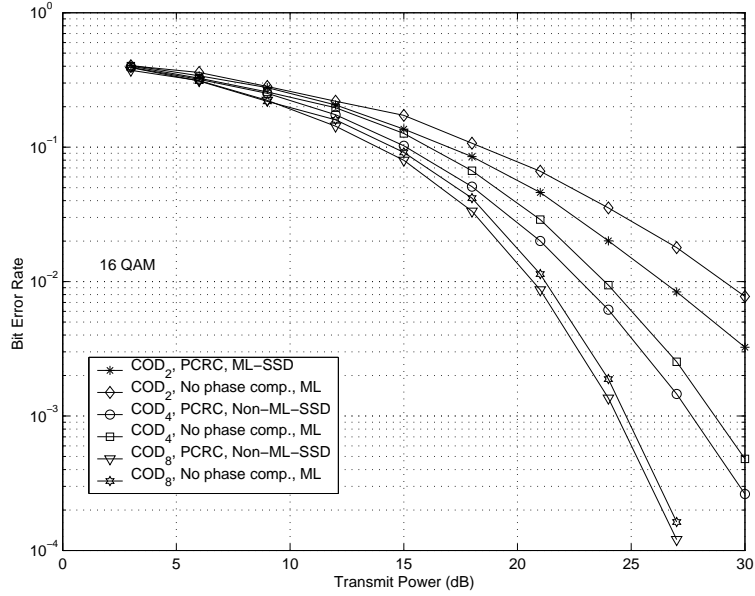}
\caption{Comparison of BER performance of $COD_2$, $COD_4$, and $COD_8$
without and with phase compensation at the relays. 16-QAM.} 
\label{fig3}
\end{center}
\end{figure}

In Fig. \ref{fig3}, we plot the BER performance for $COD_2$, $COD_4$, and 
$COD_8$ without and with phase compensation at the relays (i.e., PCRC) for 
16-QAM. Note that $COD_2$ is SSD for PCRC whereas $COD_4$ and $COD_8$
are not SSD for PCRC. So decoding of $COD_2$ with PCRC is ML-SSD, whereas 
decoding of $COD_4$ and $COD_8$ with PCRC is non-ML-SSD. When no phase 
compensation is done at the relays, we do ML decoding for all $COD_2$, 
$COD_4$, and $COD_8$. The following observations can be made from Fig. 
\ref{fig3}: $i)$ $COD_2$ without and with phase compensation at the
relays (PCRC) achieve the full diversity order of 2, $ii)$ $COD_2$ with 
PCRC and ML-SSD achieves better performance by about 3 dB at a BER of 
$10^{-2}$ compared to ML decoding of $COD_2$ without phase compensation, 
and $iii)$ even the non-ML-SSD of $COD_4$ and $COD_8$ with PCRC achieves 
full diversity of 4 and 8, respectively (but not the ML performance 
corresponding to PCRC), and even with this suboptimum decoding, PCRC 
achieves about 1 dB and 0.5 dB better performance at a BER of $10^{-2}$,
respectively, compared to ML decoding of $COD_4$ and $COD_8$ without 
phase compensation at the relays.

\begin{figure}
\begin{center}
\epsfxsize=10.0cm
\epsfbox{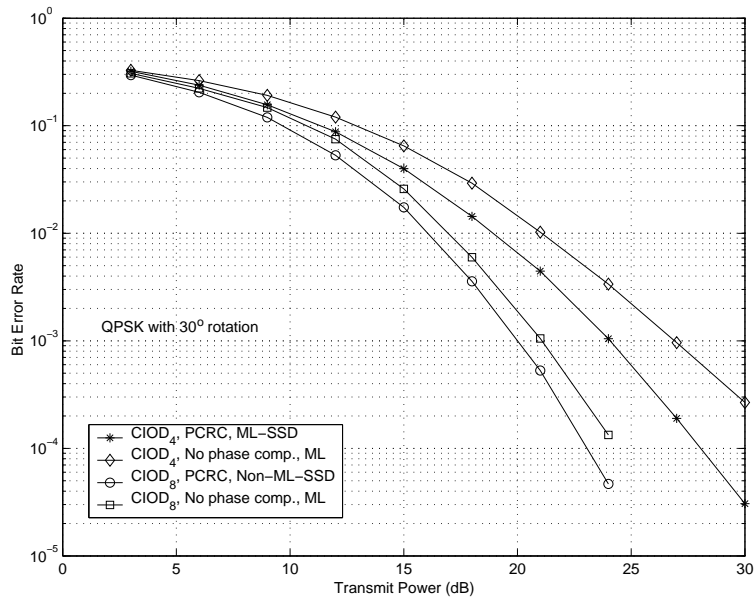}
\caption{Comparison of BER performance of $CIOD_4$ and $CIOD_8$ 
without and with with phase compensation at the relays. QPSK with 
$30^o$ rotation.} 
\label{fig4}
\end{center}
\end{figure}

In Fig. \ref{fig4}, we present a similar BER performance comparison for 
CIODs without and with phase compensation at the relays. QPSK modulation
with $30^\circ$ rotation of the constellation is used. Here again,
both $CIOD_4$ and $CIOD_8$ achieve their full diversities of 4 and 8,
respectively. We further observe that $CIOD_4$ (which is SSD for PCRC) 
with PCRC and ML-SSD achieves better performance by about 3 dB at a BER 
of $10^{-3}$ compared to ML decoding of $CIOD_4$ without phase compensation. 
Likewise, $CIOD_8$ (which is not SSD for PCRC) with PCRC and non-ML-SSD 
achieves better performance by about 1 dB at a BER of $10^{-3}$ compared 
to ML decoding of $CIOD_8$ without phase compensation.

\begin{figure}
\begin{center}
\epsfxsize=10.0cm
\epsfbox{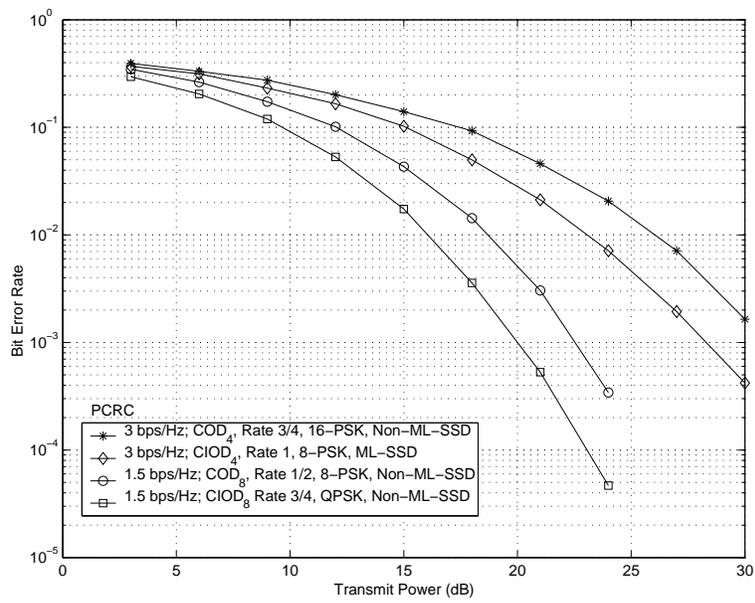}
\caption{Comparison of BER performance of CODs and CIODs with phase
compensation at the relays (i.e., PCRC) for a given spectral efficiency: 
$i)$ 3 bps/Hz; rate-3/4 $COD_4$ with 16-PSK versus rate-1 $CIOD_4$ with 
8-PSK ($10^\circ$ rotation), and $ii)$ 1.5 bps/Hz; rate-1/2 $COD_8$ with
8-PSK versus rate-3/4 $CIOD_8$ with QPSK ($30^\circ$ rotation). }
\label{fig5}
\end{center}
\end{figure}

Finally, a performance comparison between CODs and CIODs with PCRC 
for a given spectral efficiency is presented in Fig. \ref{fig5}.
A comparison at a spectral efficiency of 3 bps/Hz is made between 
$i)$ $COD_4$ with rate-3/4 and 16-PSK (spectral efficiency = 
$\frac{3}{4} \times \log_2 16 = 3$ bps/Hz), and $ii)$ $CIOD_4$ with rate-1 
and 8-PSK with $10^\circ$ rotation (spectral efficiency = 
$1\times \log_2 8 = 3$ bps/Hz). Likewise, a comparison is made at
a spectral efficiency of 1.5 bps/Hz between $COD_8$ and $CIOD_8$.  
It can be observed that, as in the case of co-located MIMO \cite{KhR}, 
in distributed STBCs with PCRC also, CIODs perform better than COD, i.e.,
coordinate interleaving improves performance. All these simulation
results reinforce the claims made in the paper in Sec. \ref{sec1}.


\begin{eqnarray}
ROD_{10} = \left [ 
\begin{array} {cccccccccc}
x_{1} &x_{9} &x_{17} &x_{18} &x_{19} &x_{20} &x_{21} &x_{22} &x_{23} &x_{24}\\ 
x_{2} &x_{10} &x_{18} &-x_{17} &x_{20} &-x_{19} &x_{22} &-x_{21} &-x_{24} &x_{23}\\ 
x_{3} &x_{11} &x_{19} &-x_{20} &-x_{17} &x_{18} &-x_{23} &-x_{24} &x_{21} &x_{22}\\ 
x_{4} &x_{12} &x_{20} &x_{19} &-x_{18} &-x_{17} &-x_{24} &x_{23} &-x_{22} &x_{21}\\ 
x_{5} &x_{13} &x_{21} &-x_{22} &x_{23} &x_{24} &-x_{17} &x_{18} &-x_{19} &-x_{20}\\ 
x_{6} &x_{14} &x_{22} &x_{21} &x_{24} &-x_{23} &-x_{18} &-x_{17} &x_{20} &-x_{19}\\ 
x_{7} &x_{15} &x_{23} &x_{24} &-x_{21} &x_{22} &x_{19} &-x_{20} &-x_{17} &-x_{18}\\ 
x_{8} &x_{16} &x_{24} &-x_{23} &-x_{22} &-x_{21} &x_{20} &x_{19} &x_{18} &-x_{17}\\ 
x_{9} &-x_{1} &x_{25} &x_{26} &x_{27} &x_{28} &x_{29} &x_{30} &x_{31} &x_{32}\\ 
x_{10} &-x_{2} &x_{26} &-x_{25} &x_{28} &-x_{27} &x_{30} &-x_{29} &-x_{32} &x_{31}\\ 
x_{11} &-x_{3} &x_{27} &-x_{28} &-x_{25} &x_{26} &-x_{31} &-x_{32} &x_{29} &x_{30}\\ 
x_{12} &-x_{4} &x_{28} &x_{27} &-x_{26} &-x_{25} &-x_{32} &x_{31} &-x_{30} &x_{29}\\ 
x_{13} &-x_{5} &x_{29} &-x_{30} &x_{31} &x_{32} &-x_{25} &x_{26} &-x_{27} &-x_{28}\\ 
x_{14} &-x_{6} &x_{30} &x_{29} &x_{32} &-x_{31} &-x_{26} &-x_{25} &x_{28} &-x_{27}\\ 
x_{15} &-x_{7} &x_{31} &x_{32} &-x_{29} &x_{30} &x_{27} &-x_{28} &-x_{25} &-x_{26}\\ 
x_{16} &-x_{8} &x_{32} &-x_{31} &-x_{30} &-x_{29} &x_{28} &x_{27} &x_{26} &-x_{25}\\ 
-x_{17} &-x_{25} &x_{1} &x_{2} &x_{3} &x_{4} &x_{5} &x_{6} &x_{7} &x_{8}\\ 
-x_{18} &-x_{26} &x_{2} &-x_{1} &x_{4} &-x_{3} &x_{6} &-x_{5} &-x_{8} &x_{7}\\ 
-x_{19} &-x_{27} &x_{3} &-x_{4} &-x_{1} &x_{2} &-x_{7} &-x_{8} &x_{5} &x_{6}\\ 
-x_{20} &-x_{28} &x_{4} &x_{3} &-x_{2} &-x_{1} &-x_{8} &x_{7} &-x_{6} &x_{5}\\ 
-x_{21} &-x_{29} &x_{5} &-x_{6} &x_{7} &x_{8} &-x_{1} &x_{2} &-x_{3} &-x_{4}\\ 
-x_{22} &-x_{30} &x_{6} &x_{5} &x_{8} &-x_{7} &-x_{2} &-x_{1} &x_{4} &-x_{3}\\ 
-x_{23} &-x_{31} &x_{7} &x_{8} &-x_{5} &x_{6} &x_{3} &-x_{4} &-x_{1} &-x_{2}\\ 
-x_{24} &-x_{32} &x_{8} &-x_{7} &-x_{6} &-x_{5} &x_{4} &x_{3} &x_{2} &-x_{1}\\ 
-x_{25} &x_{17} &x_{9} &x_{10} &x_{11} &x_{12} &x_{13} &x_{14} &x_{15} &x_{16}\\ 
-x_{26} &x_{18} &x_{10} &-x_{9} &x_{12} &-x_{11} &x_{14} &-x_{13} &-x_{16} &x_{15}\\ 
-x_{27} &x_{19} &x_{11} &-x_{12} &-x_{9} &x_{10} &-x_{15} &-x_{16} &x_{13} &x_{14}\\ 
-x_{28} &x_{20} &x_{12} &x_{11} &-x_{10} &-x_{9} &-x_{16} &x_{15} &-x_{14} &x_{13}\\ 
-x_{29} &x_{21} &x_{13} &-x_{14} &x_{15} &x_{16} &-x_{9} &x_{10} &-x_{11} &-x_{12}\\ 
-x_{30} &x_{22} &x_{14} &x_{13} &x_{16} &-x_{15} &-x_{10} &-x_{9} &x_{12} &-x_{11}\\ 
-x_{31} &x_{23} &x_{15} &x_{16} &-x_{13} &x_{14} &x_{11} &-x_{12} &-x_{9} &-x_{10}\\ 
-x_{32} &x_{24} &x_{16} &-x_{15} &-x_{14} &-x_{13} &x_{12} &x_{11} &x_{10} &-x_{9}\\ 
\end{array} \right ] 
\label{rod10}
\end{eqnarray}

\newpage

\begin{scriptsize} 
\begin{eqnarray}
ROD_{12} = \left [ 
\begin{array} {cccccccccccc}
x_{1} &x_{9} &x_{17} &x_{25} &x_{33} &x_{34} &x_{35} &x_{36} &x_{37} &x_{38} &x_{39} &x_{40}\\ 
x_{2} &x_{10} &x_{18} &x_{26} &x_{34} &-x_{33} &x_{36} &-x_{35} &x_{38} &-x_{37} &-x_{40} &x_{39}\\ 
x_{3} &x_{11} &x_{19} &x_{27} &x_{35} &-x_{36} &-x_{33} &x_{34} &-x_{39} &-x_{40} &x_{37} &x_{38}\\ 
x_{4} &x_{12} &x_{20} &x_{28} &x_{36} &x_{35} &-x_{34} &-x_{33} &-x_{40} &x_{39} &-x_{38} &x_{37}\\ 
x_{5} &x_{13} &x_{21} &x_{29} &x_{37} &-x_{38} &x_{39} &x_{40} &-x_{33} &x_{34} &-x_{35} &-x_{36}\\ 
x_{6} &x_{14} &x_{22} &x_{30} &x_{38} &x_{37} &x_{40} &-x_{39} &-x_{34} &-x_{33} &x_{36} &-x_{35}\\ 
x_{7} &x_{15} &x_{23} &x_{31} &x_{39} &x_{40} &-x_{37} &x_{38} &x_{35} &-x_{36} &-x_{33} &-x_{34}\\ 
x_{8} &x_{16} &x_{24} &x_{32} &x_{40} &-x_{39} &-x_{38} &-x_{37} &x_{36} &x_{35} &x_{34} &-x_{33}\\ 
x_{9} &-x_{1} &x_{25} &-x_{17} &x_{41} &x_{42} &x_{43} &x_{44} &x_{45} &x_{46} &x_{47} &x_{48}\\ 
x_{10} &-x_{2} &x_{26} &-x_{18} &x_{42} &-x_{41} &x_{44} &-x_{43} &x_{46} &-x_{45} &-x_{48} &x_{47}\\ 
x_{11} &-x_{3} &x_{27} &-x_{19} &x_{43} &-x_{44} &-x_{41} &x_{42} &-x_{47} &-x_{48} &x_{45} &x_{46}\\ 
x_{12} &-x_{4} &x_{28} &-x_{20} &x_{44} &x_{43} &-x_{42} &-x_{41} &-x_{48} &x_{47} &-x_{46} &x_{45}\\ 
x_{13} &-x_{5} &x_{29} &-x_{21} &x_{45} &-x_{46} &x_{47} &x_{48} &-x_{41} &x_{42} &-x_{43} &-x_{44}\\ 
x_{14} &-x_{6} &x_{30} &-x_{22} &x_{46} &x_{45} &x_{48} &-x_{47} &-x_{42} &-x_{41} &x_{44} &-x_{43}\\ 
x_{15} &-x_{7} &x_{31} &-x_{23} &x_{47} &x_{48} &-x_{45} &x_{46} &x_{43} &-x_{44} &-x_{41} &-x_{42}\\ 
x_{16} &-x_{8} &x_{32} &-x_{24} &x_{48} &-x_{47} &-x_{46} &-x_{45} &x_{44} &x_{43} &x_{42} &-x_{41}\\ 
x_{17} &-x_{25} &-x_{1} &x_{9} &x_{49} &x_{50} &x_{51} &x_{52} &x_{53} &x_{54} &x_{55} &x_{56}\\ 
x_{18} &-x_{26} &-x_{2} &x_{10} &x_{50} &-x_{49} &x_{52} &-x_{51} &x_{54} &-x_{53} &-x_{56} &x_{55}\\ 
x_{19} &-x_{27} &-x_{3} &x_{11} &x_{51} &-x_{52} &-x_{49} &x_{50} &-x_{55} &-x_{56} &x_{53} &x_{54}\\ 
x_{20} &-x_{28} &-x_{4} &x_{12} &x_{52} &x_{51} &-x_{50} &-x_{49} &-x_{56} &x_{55} &-x_{54} &x_{53}\\ 
x_{21} &-x_{29} &-x_{5} &x_{13} &x_{53} &-x_{54} &x_{55} &x_{56} &-x_{49} &x_{50} &-x_{51} &-x_{52}\\ 
x_{22} &-x_{30} &-x_{6} &x_{14} &x_{54} &x_{53} &x_{56} &-x_{55} &-x_{50} &-x_{49} &x_{52} &-x_{51}\\ 
x_{23} &-x_{31} &-x_{7} &x_{15} &x_{55} &x_{56} &-x_{53} &x_{54} &x_{51} &-x_{52} &-x_{49} &-x_{50}\\ 
x_{24} &-x_{32} &-x_{8} &x_{16} &x_{56} &-x_{55} &-x_{54} &-x_{53} &x_{52} &x_{51} &x_{50} &-x_{49}\\ 
x_{25} &x_{17} &-x_{9} &-x_{1} &x_{57} &x_{58} &x_{59} &x_{60} &x_{61} &x_{62} &x_{63} &x_{64}\\ 
x_{26} &x_{18} &-x_{10} &-x_{2} &x_{58} &-x_{57} &x_{60} &-x_{59} &x_{62} &-x_{61} &-x_{64} &x_{63}\\ 
x_{27} &x_{19} &-x_{11} &-x_{3} &x_{59} &-x_{60} &-x_{57} &x_{58} &-x_{63} &-x_{64} &x_{61} &x_{62}\\ 
x_{28} &x_{20} &-x_{12} &-x_{4} &x_{60} &x_{59} &-x_{58} &-x_{57} &-x_{64} &x_{63} &-x_{62} &x_{61}\\ 
x_{29} &x_{21} &-x_{13} &-x_{5} &x_{61} &-x_{62} &x_{63} &x_{64} &-x_{57} &x_{58} &-x_{59} &-x_{60}\\ 
x_{30} &x_{22} &-x_{14} &-x_{6} &x_{62} &x_{61} &x_{64} &-x_{63} &-x_{58} &-x_{57} &x_{60} &-x_{59}\\ 
x_{31} &x_{23} &-x_{15} &-x_{7} &x_{63} &x_{64} &-x_{61} &x_{62} &x_{59} &-x_{60} &-x_{57} &-x_{58}\\ 
x_{32} &x_{24} &-x_{16} &-x_{8} &x_{64} &-x_{63} &-x_{62} &-x_{61} &x_{60} &x_{59} &x_{58} &-x_{57}\\ 
-x_{33} &-x_{41} &-x_{49} &-x_{57} &x_{1} &x_{2} &x_{3} &x_{4} &x_{5} &x_{6} &x_{7} &x_{8}\\ 
-x_{34} &-x_{42} &-x_{50} &-x_{58} &x_{2} &-x_{1} &x_{4} &-x_{3} &x_{6} &-x_{5} &-x_{8} &x_{7}\\ 
-x_{35} &-x_{43} &-x_{51} &-x_{59} &x_{3} &-x_{4} &-x_{1} &x_{2} &-x_{7} &-x_{8} &x_{5} &x_{6}\\ 
-x_{36} &-x_{44} &-x_{52} &-x_{60} &x_{4} &x_{3} &-x_{2} &-x_{1} &-x_{8} &x_{7} &-x_{6} &x_{5}\\ 
-x_{37} &-x_{45} &-x_{53} &-x_{61} &x_{5} &-x_{6} &x_{7} &x_{8} &-x_{1} &x_{2} &-x_{3} &-x_{4}\\ 
-x_{38} &-x_{46} &-x_{54} &-x_{62} &x_{6} &x_{5} &x_{8} &-x_{7} &-x_{2} &-x_{1} &x_{4} &-x_{3}\\ 
-x_{39} &-x_{47} &-x_{55} &-x_{63} &x_{7} &x_{8} &-x_{5} &x_{6} &x_{3} &-x_{4} &-x_{1} &-x_{2}\\ 
-x_{40} &-x_{48} &-x_{56} &-x_{64} &x_{8} &-x_{7} &-x_{6} &-x_{5} &x_{4} &x_{3} &x_{2} &-x_{1}\\ 
-x_{41} &x_{33} &-x_{57} &x_{49} &x_{9} &x_{10} &x_{11} &x_{12} &x_{13} &x_{14} &x_{15} &x_{16}\\ 
-x_{42} &x_{34} &-x_{58} &x_{50} &x_{10} &-x_{9} &x_{12} &-x_{11} &x_{14} &-x_{13} &-x_{16} &x_{15}\\ 
-x_{43} &x_{35} &-x_{59} &x_{51} &x_{11} &-x_{12} &-x_{9} &x_{10} &-x_{15} &-x_{16} &x_{13} &x_{14}\\ 
-x_{44} &x_{36} &-x_{60} &x_{52} &x_{12} &x_{11} &-x_{10} &-x_{9} &-x_{16} &x_{15} &-x_{14} &x_{13}\\ 
-x_{45} &x_{37} &-x_{61} &x_{53} &x_{13} &-x_{14} &x_{15} &x_{16} &-x_{9} &x_{10} &-x_{11} &-x_{12}\\ 
-x_{46} &x_{38} &-x_{62} &x_{54} &x_{14} &x_{13} &x_{16} &-x_{15} &-x_{10} &-x_{9} &x_{12} &-x_{11}\\ 
-x_{47} &x_{39} &-x_{63} &x_{55} &x_{15} &x_{16} &-x_{13} &x_{14} &x_{11} &-x_{12} &-x_{9} &-x_{10}\\ 
-x_{48} &x_{40} &-x_{64} &x_{56} &x_{16} &-x_{15} &-x_{14} &-x_{13} &x_{12} &x_{11} &x_{10} &-x_{9}\\ 
-x_{49} &x_{57} &x_{33} &-x_{41} &x_{17} &x_{18} &x_{19} &x_{20} &x_{21} &x_{22} &x_{23} &x_{24}\\ 
-x_{50} &x_{58} &x_{34} &-x_{42} &x_{18} &-x_{17} &x_{20} &-x_{19} &x_{22} &-x_{21} &-x_{24} &x_{23}\\ 
-x_{51} &x_{59} &x_{35} &-x_{43} &x_{19} &-x_{20} &-x_{17} &x_{18} &-x_{23} &-x_{24} &x_{21} &x_{22}\\ 
-x_{52} &x_{60} &x_{36} &-x_{44} &x_{20} &x_{19} &-x_{18} &-x_{17} &-x_{24} &x_{23} &-x_{22} &x_{21}\\ 
-x_{53} &x_{61} &x_{37} &-x_{45} &x_{21} &-x_{22} &x_{23} &x_{24} &-x_{17} &x_{18} &-x_{19} &-x_{20}\\ 
-x_{54} &x_{62} &x_{38} &-x_{46} &x_{22} &x_{21} &x_{24} &-x_{23} &-x_{18} &-x_{17} &x_{20} &-x_{19}\\ 
-x_{55} &x_{63} &x_{39} &-x_{47} &x_{23} &x_{24} &-x_{21} &x_{22} &x_{19} &-x_{20} &-x_{17} &-x_{18}\\ 
-x_{56} &x_{64} &x_{40} &-x_{48} &x_{24} &-x_{23} &-x_{22} &-x_{21} &x_{20} &x_{19} &x_{18} &-x_{17}\\ 
-x_{57} &-x_{49} &x_{41} &x_{33} &x_{25} &x_{26} &x_{27} &x_{28} &x_{29} &x_{30} &x_{31} &x_{32}\\ 
-x_{58} &-x_{50} &x_{42} &x_{34} &x_{26} &-x_{25} &x_{28} &-x_{27} &x_{30} &-x_{29} &-x_{32} &x_{31}\\ 
-x_{59} &-x_{51} &x_{43} &x_{35} &x_{27} &-x_{28} &-x_{25} &x_{26} &-x_{31} &-x_{32} &x_{29} &x_{30}\\ 
-x_{60} &-x_{52} &x_{44} &x_{36} &x_{28} &x_{27} &-x_{26} &-x_{25} &-x_{32} &x_{31} &-x_{30} &x_{29}\\ 
-x_{61} &-x_{53} &x_{45} &x_{37} &x_{29} &-x_{30} &x_{31} &x_{32} &-x_{25} &x_{26} &-x_{27} &-x_{28}\\ 
-x_{62} &-x_{54} &x_{46} &x_{38} &x_{30} &x_{29} &x_{32} &-x_{31} &-x_{26} &-x_{25} &x_{28} &-x_{27}\\ 
-x_{63} &-x_{55} &x_{47} &x_{39} &x_{31} &x_{32} &-x_{29} &x_{30} &x_{27} &-x_{28} &-x_{25} &-x_{26}\\ 
-x_{64} &-x_{56} &x_{48} &x_{40} &x_{32} &-x_{31} &-x_{30} &-x_{29} &x_{28} &x_{27} &x_{26} &-x_{25}\\ 
\end{array} \right ] 
\label{rod12}
\end{eqnarray}
\end{scriptsize} 

\section{Conclusions}
\label{sec6}
We summarize the conclusions in this paper and future work as follows.
Amplify-and-forward (AF) schemes in cooperative communications are
attractive because of their simplicity. Full diversity (FD),
linear-complexity single symbol decoding (SSD), and high rates of
DSTBCs are three important attributes to work towards  AF cooperative
communications. Earlier work in \cite{YiK} has shown that, without
assuming phase knowledge at the relays, FD and SSD can be achieved
in AF distributed orthogonal STBC schemes; however, the rate achieved
decreases linearly with the number of relays $N$. Our work in this paper
established that if phase knowledge is exploited at the relays in the
way we have proposed, then FD, SSD, and high rate can be achieved
simultaneously; in particular, the rate achieved in our scheme can be
$\frac{1}{2}$, which is independent of the number of relays $N$. We
proved the SSD for our scheme in Theorem 2. FD was proved in Theorem 6.
Rate-1/2 construction for any $N$ was presented in Theorem 5. In addition
to these results, we also established other results regarding $i)$
invariance of SSD under coordinate interleaving (Theorem 4), and $ii)$
retention of FD even with single-symbol non-ML decoding. Simulation results
confirming the claims were presented. All these important results have not
been shown in the literature so far. These results offer useful insights
and knowledge for the designers of future cooperative communication based
systems (e.g., cooperative communication ideas are being considered in
future evolution of standards like IEEE 802.16).

In this work, we have assumed only phase knowledge at the relays. Of course,
one can assume that both amplitude as well as the phase of source-to-relay
are known at the relay. A natural question that can arise then is `what can
amplitude knowledge at the relay (in addition to phase knowledge) buy?'
Since we have shown that phase knowledge alone is adequate to achieve FD,
some extra coding gain may be possible with amplitude knowledge. This
aspect of the problem is beyond the scope of this paper; but it is a valid
topic for future work.

{\footnotesize 
\bibliographystyle{IEEE}

}

\newpage

\begin{table*}
\begin{center}
\begin{tabular}{|c|c|c|c|c|}
\hline 
Code & Number of & Rate & Necessary and sufficient & Sufficient \\
& Relays &   & Conditions (\ref{singlesymx}), (\ref{ssddstbc1}) \& (\ref{ssddstbc2} ) & Condition in \cite{RaR3} \\ \hline
$COD_2$ (Alamouti)   & $N =2$ &  1 & True & True \\ \hline
$COD_4$ & $N =4$ & 3/4  & False & False \\ \hline
$CIOD_4$& $N =4$ & 1  & True & False \\ \hline
$CUW_4$ & $N =4$ & 1   & True & True \\ \hline
$COD_8$ & $N =8$ & 1/2  & False & False \\ \hline
$CIOD_8$& $N =8$ & 3/4  & False & False \\ \hline
$RR_8$ & $N =8$ &  1  & True & False \\ \hline
CODs from RODs  & $N \geq 4$ & 1/2 &True & False \\ \hline
\end{tabular}
\caption{Test for necessary and sufficient conditions for various classes 
of codes for PCRC.}
\label{tab1}
\end{center}
\end{table*}

\end{document}